\documentclass[
    aps,prb,
    twocolumn,
    10pt,
    superscriptaddress,
    nofootinbib,
    ]{revtex4-2}

\usepackage{graphicx}
\usepackage{subfigure}
\graphicspath{{pics/}}
\usepackage{amsmath}
\usepackage[Symbol]{upgreek}
\usepackage{bm} 
\usepackage[dvipsnames]{xcolor}
\usepackage{microtype}
\usepackage{multirow}
\usepackage{lipsum}

\usepackage{siunitx}
\usepackage[
    colorlinks=true,
    allcolors=MidnightBlue,
    ]{hyperref}

\usepackage{cleveref}
\crefname{figure}{Fig.}{Figs.}
\crefname{table}{Table}{Tables}
\crefname{equation}{Eq.}{Eqs.}
\Crefname{figure}{Figure}{Figures}
\Crefname{table}{Table}{Tables}
\Crefname{equation}{Equation}{Equations}

\usepackage[shortcuts]{extdash}

\let\originalleft\left
\let\originalright\right
\renewcommand{\left}{\mathopen{}\mathclose\bgroup\originalleft}
\renewcommand{\right}{\aftergroup\egroup\originalright}

{\color{Tomette}}{\ignorespacesafterend}
{\color{MidDarkAqua}}{\ignorespacesafterend}
{\color{black}}{\ignorespacesafterend}

\renewcommand{\vec}[1]{\mathbf{#1}}
\renewcommand{\Re}{\mathrm{Re}}
\renewcommand{\Im}{\mathrm{Im}}
\newcommand*{\vG}{\vec{G}}
\newcommand*{\vk}{\vec{k}}
\newcommand*{\vq}{\vec{q}}
\newcommand*{\Vdft}{$\Omega_\mathrm{DFT}$}
\newcommand*{\Vexp}{$\Omega_\mathrm{295K}$}

\newcommand*{\intd}[1]{\mathrm{d}#1\,}
\newcommand*{\twofourtwo}{$2\times4\times2$}
\newcommand*{\foursixfour}{$4\times6\times4$}
\newcommand*{\sixeightsix}{$6\times8\times6$}

\newcommand*{\twelvefourteentwelve}{$12\times14\times12$}

\begin{document}


\title{Band gap renormalization, carrier mobilities, and the electron-phonon self-energy in crystalline naphthalene}


\author{Florian Brown-Altvater}
\email[]{altvater@berkeley.edu}
\affiliation{Department of Chemistry, University of California, Berkeley, 
  California, USA}
\affiliation{Molecular Foundry, Lawrence Berkeley National Laboratory,
  Berkeley, California, USA}

\author{Gabriel Antonius}
\affiliation{D\'epartement de Chimie, Biochimie et Physique, Institut de Recherche sur l'Hydrog\`ene, Universit\'e du Qu\'ebec \`a Trois-Rivi\`eres, C.P. 500, Trois-Rivi\`eres,  Canada}
\affiliation{Department of Physics, University of California, Berkeley, 
  California, USA}

\author{Tonatiuh Rangel}
\affiliation{Molecular Foundry, Lawrence Berkeley National Laboratory, 
  Berkeley, California, USA}
\affiliation{Department of Physics, University of California, Berkeley, 
  California, USA}

\author{Matteo Giantomassi}
\affiliation{Universit\'e catholique de Louvain, Louvain-la-Neuve, Belgium}

\author{Claudia Draxl}
\affiliation{Humboldt Universit\"at Berlin, Germany}

\author{Xavier Gonze}
\affiliation{Universit\'e catholique de Louvain, Louvain-la-neuve, Belgium}
\affiliation{Skolkovo Institute of Science and Technology, Moscow, Russia}
  
\author{Steven G. Louie}
\affiliation{Department of Physics, University of California, Berkeley, 
  California, USA}
\affiliation{Materials Sciences Division, Lawrence Berkeley National Laboratory, 
  Berkeley, California, USA}

\author{Jeffrey B. Neaton}
\email[]{jbneaton@berkeley.edu}
\affiliation{Molecular Foundry, Lawrence Berkeley National Laboratory, 
  Berkeley, California, USA}
\affiliation{Department of Physics, University of California, Berkeley, 
  California, USA}
\affiliation{Kavli Energy NanoSciences Institute at Berkeley, California, USA}

\date{\today}

\begin{abstract}

Organic molecular crystals are expected to feature appreciable electron-phonon interactions
  that influence their electronic properties
  at zero and finite temperature.
In this work, we report first-principles
  calculations and an analysis of
  the electron-phonon self-energy in naphthalene crystals.
We compute
  the zero-point renormalization and temperature dependence
  of the fundamental band gap, and the resulting scattering lifetimes
  of electronic states near the valence- and conduction-band edges
  employing density functional theory.
Further, our calculated phonon renormalization of the $GW$-corrected quasiparticle band structure
  predicts a fundamental band gap of \SI{5}{eV} for naphthalene at room temperature,
  in good agreement with experiments. 
From our calculated phonon-induced electron lifetimes,
  we obtain the temperature-dependent mobilities
  of electrons and holes in good agreement
  with experimental measurements at room temperatures.
Finally, we show that an approximate energy self-consistent computational scheme
  for the electron-phonon self-energy leads to the prediction of
  strong satellite bands in the electronic band structure.
We find
  that a single calculation of the self-energy
  can reproduce the self-consistent results of the band gap renormalization and electrical mobilities for naphthalene,
  provided that the on-the-mass-shell approximation is used, i.e., if the self-energy is evaluated at the bare eigenvalues.

\end{abstract}
  

\maketitle


Molecular crystals, periodic arrays of molecules bound by noncovalent interactions, can
  nonetheless feature 
  relatively high charge carrier mobilities~\cite{Jurchescu2004,Takeya2007,Minemawari2011,Fratini2017}.
The acene family of molecular crystals are of particular interest,
  having high crystalline purity,
  making them attractive for fundamental studies and 
  various optoelectronic applications~%
  \cite{Dediu2009,Ortmann2010,Gao2015,Wang2018}.
In acenes, each monomer consists of a rigid unit of fused benzene rings.
These monomers crystallize in a herringbone structure (\cref{fig:naph_structure}).
Naphthalene, the second smallest of the acene family,
  provides a popular testbed for electronic structure calculations and experiments,
  with results that can often be extrapolated to its larger siblings~%
  \cite{Hannewald2004b}.

Electron-phonon coupling (EPC)
  has long been understood to be important in determining
  the electronic and transport properties
  of these materials \cite{Pope1999,Ortmann2011,Ostroverkhova2016}.
Along with contributions from thermal lattice expansion,
  the EPC is responsible for the temperature-dependent renormalization
  of the band structure.
Electron-phonon scattering and decay channels also result in finite lifetimes
  for electronic states and limit charge carrier mobilities.
The finite lifetimes result in a broadening of the electronic bands that can be observed with
  photoemission spectroscopy, for example \cite{Ohtomo2009,Hatch2010}.

The vast majority of prior theoretical studies of temperature effects in organic crystals arising from EPC
  focus on lifetimes
  and mobilities of charge carriers~%
\cite{Hannewald2004,Troisi2009,Girlando2010,Northrup2011,Casula2012,Xi2012,Kobayashi2013,Heck2015,Ostroverkhova2016,Fratini2017,Ishii2017,Oberhofer2017,Lee2018,Stehr2016,Fratini2017,Ishii2018}.
Prior \textit{ab initio} studies that explicitly calculate the renormalization of band gaps are usually limited to few-atom systems~\cite{Marini2008,Giustino2010,Kawai2014,Ponce2015a,Nery2016,Antonius2014} or small molecules~\cite{Monserrat2015}.
One study that calculated both the broadening and renormalization of the band gap of pentacene crystals
  used a tight-binding model parametrized by many-body perturbation theory (MBPT) calculations~\cite{Ciuchi2012}, reporting unusual quasi-discontinuities in the band structure caused by EPC that have been corroborated by experimental results, showing ``kinks'' in the electronic dispersion~\cite{Ciuchi2012,Bussolotti2017}.
In another study, \citet{Vukmirovic2012} evaluated the EPC matrix elements for two pairs of bands in naphthalene using first-principles methods.
They reported weak EPC, strengthening the argument for band-like charge carrier transport.
\Citet{Lee2018} use a fully \textit{ab initio} approach to calculate the temperature-dependent hole mobility.

In this work, we compute from first principles the temperature dependence
  of the band structure and the electron and hole transport properties of naphthalene crystals.
We use density functional theory and the dynamical Allen-Heine-Cardona theory to compute
  both the real and imaginary contributions to the electron-phonon self-energy.
With this quantity, we predict the temperature renormalization of the band gap,
  and we obtain the hole and electron mobilities
  within the relaxation-time approximation.
We discuss the details of the calculated frequency-dependent electron-phonon self-energy of the electron or hole,
  and identify features that should apply to acene and other molecular solids,
  such as the approximate independence of the self-energy on the electron wave vector $\vk$.
We find that in naphthalene, the band dispersion, phonon frequencies,
  and the renormalization energies are of the same order of magnitude,
  challenging the validity of perturbation theory in this system.
We address this issue by exploring a self-consistent computational scheme for the electron-phonon self-energy,
  and we show that a single calculation of the self-energy
  can reproduce self-consistent results of the band gap renormalization and charge carrier mobilities,
  provided that the on-the-mass-shell approximation is used.

\section{Theory and Methods}

\subsection{Theoretical Framework}

The starting point for our calculations is density functional theory (DFT),
  which provides Kohn-Sham orbital wave functions $\psi_{n\vk}$
  and orbital energies $\epsilon^0_{n\vk}$, where $n$ is the band index
  and $\vk$ is the wave vector.
We rely on density functional perturbation theory (DFPT)
  to compute the phonon coupling potential,
  and we incorporate the electron-phonon interactions via many-body perturbation theory,
  specifically a low-order diagrammatic expansion of the electron-phonon self-energy~\cite{Ponce2014,Marini2015,Giustino2017}.

To obtain the electron-phonon self-energy we follow the approach described in~\cite{Antonius2015,Ponce2015a}.
To lowest order in perturbation theory, the electron-phonon self-energy $\Sigma^\mathrm{ep}_{n\vk}$ can be divided into two terms, namely the Fan and Debye-Waller (DW) terms
\begin{align}
 \Sigma^\mathrm{ep}_{n\vk}(\omega, T) = 
    \Sigma^\text{Fan}_{n\vk}(\omega, T) 
    + \Sigma^\text{DW}_{n\vk}(T).
 \label{eq:sigma}
\end{align}
We briefly summarize each term. 
The frequency-dependent Fan term is given as \begin{align}
 & \Sigma^\text{Fan}_{n\vk}(\omega, T) 
  ={} 
    \sum_{\nu\vq} 
      \frac{1}{2\omega_{\nu\vq}}
      \sum_{m}
      \vert g_{nm\nu}(\vk,\vq) \vert^2
  \notag\\
  & \times 
    \bigg[
      \frac{N_{\nu\vq}(T) + f_{m\vk+\vq}(T)}
           {\omega - \varepsilon_{m\vk+\vq}^0 
            + \omega_{\nu\vq} + i\eta}
      + \frac{N_{\nu\vq}(T) + 1- f_{m\vk+\vq}(T)}
             {\omega - \varepsilon_{m\vk+\vq}^0 
              - \omega_{\nu\vq} + i\eta}
    \bigg].
\label{eq:Fan}
\end{align}
In \cref{eq:Fan}, the phonon modes are specified by indices $\nu$, wave vector $\vq$, and energies $\omega_{\nu\vq}$.
Phonons couple electrons in state $n\vk$ with state $m\vk\!+\!\vq$ through the first derivative of the electron crystal potential $V_{\nu\vq}^{(1)}$ associated with the respective phonon's atomic displacement pattern.
The electron-phonon matrix elements $g_{nm\nu}(\vk,\vq)\!=\!\langle \psi_{n\vk} \mid V_{\nu\vq}^{(1)} \mid \psi_{m\vk+\vq} \rangle$ %
 determine the coupling strength between the electronic states and the phonons.
The temperature dependence of the Fan term arises from the phonon ($N$) and electron ($f$) occupation factors.
We can see that even at zero temperature, the self-energy has a finite value.
The denominators give rise to poles at $\omega = \varepsilon^0 \pm \omega_{\nu\vq}$,
  which are rendered smooth with the parameter $\eta$;
  $\eta$, in principle, is real, infinitesimal and has the same sign as $\omega$ in \cref{eq:Fan}, which yields the time-ordered self-energy, in contrast to the retarded self-energy \cite{Giustino2017}.
In practice, we use a value of \SI{0.025}{eV} to account for the finite $\vq$-grid sampling.
Details of the convergence of the self-energy with respect to $\vq$-grid and $\eta$ can be found in the supplemental material~\cite{SM}.

The frequency independent Debye-Waller term
\begin{equation}
    \Sigma^\text{DW}_{n\vk}(T) = 
    \sum_{\nu\vq}
      \frac{1}{2\omega_{\nu\vq}}
      \langle n\vk \mid V_{\nu\vq,\nu\vq}^{(2)} \mid n\vk \rangle       \big[ 2N_{\nu\vq}(T)+1 \big]
\label{eq:DW}
\end{equation}
makes up the second part of the electron-phonon self-energy.
The DW term depends on the second derivative of the potential $V_{\nu\vq,\nu\vq}^{(2)}$, which is somewhat more arduous to calculate.
We use the rigid-ion approximation, which allows us to write \cref{eq:DW} in terms of the first derivative~\cite{Allen1976,Gonze2011,Ponce2014}.
In this way, we can obtain all values from DFT and DFPT calculations.

There are two main challenges in calculating the self-energy efficiently.
The first challenge is that $\vq$-space has to be sampled more densely
  compared to a typical phonon band structure calculation,
  which rapidly becomes the main bottleneck for large systems.
%
%
In this work, 
  we interpolate the phonon coupling potential in real space,
  following prior work~%
  \cite{Eiguren2008a,Verdi2015,Gonze2020,Sjakste2015}.
%
It is standard practice to interpolate the phonon frequencies
  of a regular $\vq$-grid onto arbitrary $\vq$-points by means of a
  Fourier transform of the dynamical matrices to real space, and back to reciprocal space.
Applying the same principle here, we calculate the potential derivative with DFPT
  on a coarse $\vq$-point grid and interpolate to a finer grid via Fourier transform.
We define the long-range component of the phonon potential of atom $\kappa$ along the Cartesian direction $j$ as 
\begin{align}
    V^{L}_{\kappa j}(\vq, \vec{r}) =
    i \frac{4\pi}{\Omega}
    \sum_{\vG\neq-\vq}
      \frac{
            e^{i (\vq+\vG)\cdot(\vec{r}-\tau_{\kappa})}
            (\vq+\vG)_{j'}\cdot Z^*_{\kappa,j' j}
            }
            {(\vq+\vG)\cdot \epsilon^{\infty}\cdot (\vq+\vG)},
\end{align}
where $\epsilon^{\infty}$ is the static
dielectric matrix
  without the lattice contribution to the screening,
  and $Z^*_{\kappa,j' j}$ is the Born effective charge tensor.
These quantities are computed from DFPT by including the response of
  the system to a macroscopic electric field.
The long-ranged component of the phonon potential
  represents the dipole potential created by displacing
  the Born effective charges of each atom,
  and becomes the dominant contribution to the potential
  in the limit $\vq\rightarrow 0$.
Next, we perform a Fourier transform of the short-range
  component of the phonon coupling potential,
  starting from the coarse $\vq$-point grid,
\begin{align}
    W_{\kappa j}(\vec{r}-\vec{R}_l) 
      &= \sum_\vq e^{i\vq\cdot \vec{R}_l}
        \left[
          V^{(1)}_{\kappa j}(\vq,\vec{r})
          - V^\mathrm{L}_{\kappa j}(\vq,\vec{r})
        \right],
\end{align}
where $W_{\kappa j}(\vec{r}-\vec{R}_l)$ represents
  the short-range component of the perturbative potential
  associated with the displacement of atom $\kappa$
  in the unit cell $l$ along the Cartesian direction $j$,
  and $\vec{r}$ is defined within the first unit cell ($\vec{R}_0\!=\!0$).
The interpolated phonon potential for an arbitrary point $\tilde \vq$ is then
\begin{align}
V^{(1)}_{\kappa j}(\vec{\tilde{q}}, \vec{r})
    &\approx \sum_l W_{\kappa j}(\vec{r}-\vec{R}_l)
            e^{-i\vec{\tilde{q}}\cdot \vec{R}_l}
    + V^\mathrm{L}_{\kappa j}(\vec{\tilde{q}},\vec{r}).
\end{align}
This interpolation scheme reproduces the electron-phonon coupling matrix elements
  with accuracy better than $1\%$, as shown in the supplemental material~\cite{SM}.
It achieves the same goal as the Wannier interpolation used in other works \cite{Sjakste2015,Giustino2007,Eiguren2008}, but avoids the computation of Wannier functions altogether.

The second challenge in the computation of the electron-phonon self-energy
  lies in the sum over electronic states $m$ in \cref{eq:Fan}, which can converge slowly with the number of bands.
We evaluate this sum explicitly using all valence bands,
  and conduction bands up to \SI{5}{eV} above the last electronic state
  for which the self-energy is computed.
Above this cut-off the sum over infinite bands is replaced by a Sternheimer equation,
    and their contribution to the self-energy is treated statically, an approximation that has been shown to be effective in prior work~\cite{Gonze2011,Antonius2015}.
Furthermore, this contribution is evaluated on the coarse $\vq$-grid,
    since the denominator of the self-energy in \cref{eq:Fan} is never small for these bands,
    and is thus a smooth function of $\vq$.

\subsection{Computational details}
DFT calculations are performed with the \textsc{ABINIT} code~\cite{Gonze2009,Gonze2016,Gonze2020}
  using Fritz-Haber-Institut norm-conserving pseudopotentials \cite{Fuchs1999},
  and setting the plane waves kinetic energy cutoff to \SI{45}{Ha}.
We use the Perdew-Burke-Ernzerhof (PBE) functional in combination with the Grimme-D3 correction~\cite{Grimme2010,Grimme2011} to account for London dispersion forces.
To obtain the electronic ground state density, we sample the Brillouin zone on a $\Gamma$-centered $\vk$-grid of \twofourtwo{}.
All electronic energies in this work are given relative to the valence band maximum.

The phonons and associated potential derivatives are calculated with DFPT, including the treatment of dispersion forces~\cite{Gonze1997,Baroni2001,Gonze2005,Troeye2016}.
A coarse $\Gamma$-centered \foursixfour{} $\vq$-grid gives well converged phonon frequencies and displacements after interpolation of the dynamical matrix,
  as shown in our previous work~\cite{Brown-Altvater2016}.
In the present work, we start from an even finer \sixeightsix{} grid, and we interpolate not only 
phonon frequencies and displacements, but also
the phonon potentials and self-energy onto a \twelvefourteentwelve{} $\vq$-grid,
  which converges the renormalization and broadening values within a few meV
  (see the supplemental material \cite{SM} for convergence studies).

\subsection{Lattice parameters}

\begin{figure}
  \includegraphics[]{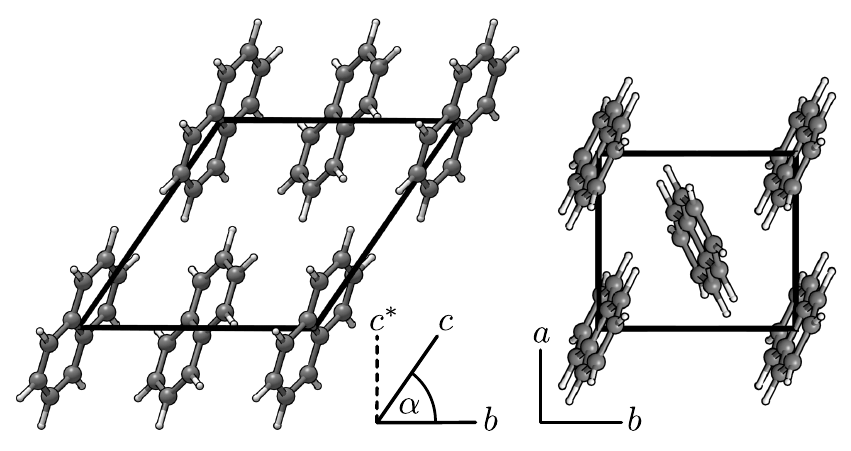}
  \caption{%
    Naphthalene is the smallest acene that crystallizes in a herringbone structure. There are two molecules in the monoclinic unit cell, each situated at inversion centers.}%
    \label{fig:naph_structure}
\end{figure}

Naphthalene crystallizes in the $P2_1/a$ space group, forming a herringbone structure with two molecules per unit cell (\cref{fig:naph_structure}) that are held together by noncovalent interactions.
As discussed in previous work~\cite{Rangel2016}, relaxing lattice parameters and atomic coordinates with van der Waals corrected functionals or pair-wise dispersion corrections results in excellent agreement with low-temperature experiments.
The relaxed unit cell volume of naphthalene obtained with PBE-D3 is within \SI{0.4}{\percent} of the experimental value measured at \SI{5}{K}%
    \footnote{The experimental crystal structures used in this work are available at the Cambridge Structural Database~\cite{Thomas2010,*csd_url}. The identifiers for the structures measured at \SI{5}{K} and \SI{295}{K} are NAPHTA31 and NAPHTA36, respectively, and published in association with~\cite{Capelli2006}.}%
.
%
We use this relaxed unit cell for most of our calculations, and we refer to it by its computed volume, \Vdft{}.

To simulate thermal lattice expansion, we use fixed experimental lattice parameters obtained at \SI{295}{K}\footnotemark[1], and we relax the internal atomic coordinates using PBE-D3.
The volume of this room-temperature structure is about \SI{6}{\percent} larger than that of the low-temperature structure.
The main expansion occurs in the $ab$ plane, and through a decreased tilt of the monoclinic cell (see the supplemental material for all unit-cell parameters \cite{SM}).
Any calculations that use this experimental lattice are labeled by this larger volume, \Vexp{}.


\section{Results and discussion}
\label{sec:results}

\subsection{Electronic and phonon band structures}

\begin{figure}
  \includegraphics[]{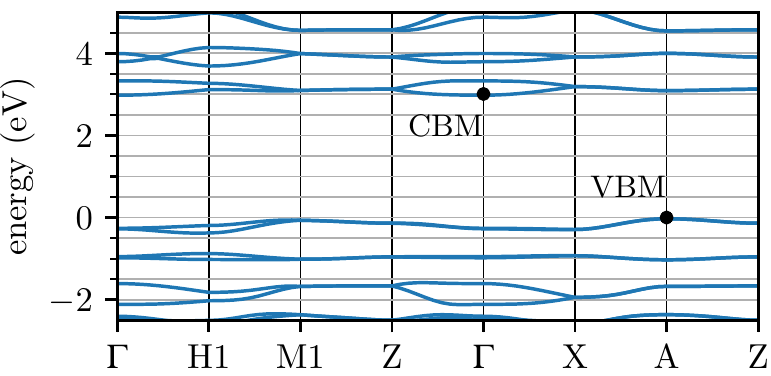}
  \caption{%
    Electronic band structure of naphthalene calculated with DFT.
    The locations of the conduction-band minimum (CBM) and valence-band maximum (VBM) are indicated with black dots.
    }\label{fig:naph_bs}
\end{figure}

The electronic band structure of naphthalene is characteristic for a small molecule crystal \cite{Rangel2016}: it possesses a sizable band gap combined with flat, well-separated groups or complexes of bands (\cref{fig:naph_bs}).
DFT yields an indirect gap of \SI{3.01}{eV} between the valence band maximum (VBM) at A and the conduction band minimum (CBM) at $\Gamma$.
The weak intermolecular interactions lead to small bandwidths for the complexes less than \SI{0.4}{eV}.
Furthermore, because naphthalene has two molecules per unit cell, the electronic bands double up in so-called Davydov pairs~\cite{Davydov1964,Sheka1975}.
In the vicinity of the band gap, these Davydov pairs are separated from each other by about \SI{0.4}{eV}.
This separation drastically reduces mixing of states from different Davydov pairs.
The wave functions of solid naphthalene at the band edges therefore vary little throughout the Brillouin zone, and closely resemble linear combinations of gas-phase-like molecular orbitals.
Dispersion and interband interactions are higher for bands just below \SI{-2}{eV} as the spacing between electronic levels decreases,
  and for bands above \SI{4.5}{eV} as the wave functions become more delocalized.

For the phonon frequencies, we obtain excellent agreement with experiments across the Brillouin zone using PBE-D3
  (see the supplemental material \cite{SM} for the full phonon band structure in comparison with experimental measurements from Refs.~\cite{Natkaniec1980,Suzuki1968}), similar to our previous results with the vdW-DF-cx functional \cite{Brown-Altvater2016}.
Since we analyzed the vibrational properties of naphthalene in depth in Ref.~\cite{Brown-Altvater2016}, we give only a brief overview of the main features here.
In naphthalene, intermolecular modes (\SI{<20}{meV}) can be clearly distinguished from intramolecular modes (\SIrange{20}{400}{meV}).
Intermolecular modes are translational and librational modes of rigid molecules, while for intramolecular modes, the phonon displacement vectors resemble linear combinations of gas phase vibrations.

We emphasize that, despite the clear separation between inter\=/~and intramolecular modes, we treat all phonon modes on the same footing in our work.
While hopping transport models often use the rigid molecule approximation~\cite{Troisi2006,Coropceanu2009,Wang2013}, it has been shown that the mixed inter- and intramolecular low-frequency modes can have large EPC contributions, especially for larger molecules like rubrene~\cite{Xie2018}.

Upon thermal lattice expansion, the spacing between molecules becomes larger.
The lowered interaction leads to softening of the intermolecular modes, decreasing the lowest frequencies by up to \SI{40}{\percent}.
In contrast, intramolecular frequencies, which depend on the covalent interatomic forces,
  are found to change very little, as shown in the supplemental material \cite{SM}.

\subsection{Temperature-dependent renormalization of the band structure}

\begin{figure}
  \includegraphics[]
    {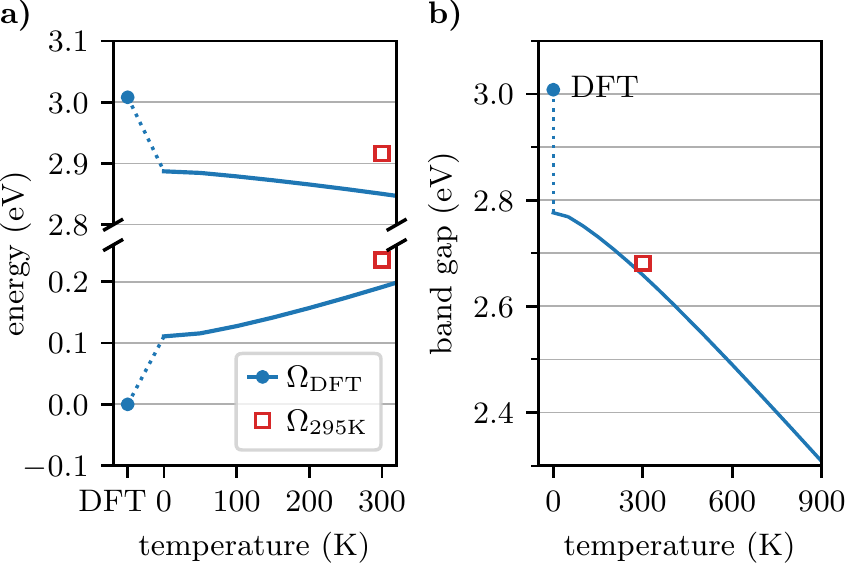}\\
  \caption{%
    a) Renormalization and temperature dependence of the band edge states at $\Gamma$ and A, with \Vdft{}.
    The dotted lines indicate the ZPR, connecting the bare eigenvalues calculated with PBD-D3 (circles) with the renormalized energies at \SI{0}{K}.
    The renormalized energies for \Vexp{} (squares) at \SI{300}{K} are plotted for comparison.
    b) ZPR (dotted) and temperature dependence (solid) of the indirect band gap of naphthalene for \Vdft{}.
    The red square shows the renormalization at \SI{300}{K} using \Vexp{}.
    }\label{fig:temp}
\end{figure}

We obtain the temperature-dependent electronic band structure of naphthalene from the real part of the electron-phonon self-energy using the on-the-mass-shell approximation~\cite{Cannuccia2012}
\begin{equation}
 \varepsilon_{n\vk}(T) =
  \varepsilon^0_{n\vk}
   + \Re\left[
        \Sigma^\mathrm{ep}_{n\vk}(\varepsilon^0_{n\vk},T)
     \right],
\label{eq:renorm}
\end{equation}
where $\varepsilon^0_{n\vk}$ is the bare DFT eigenvalue with band index $n$ and wave vector $\vk$, and $\varepsilon_{n\vk}$ is the renormalized energy.

The temperature dependence of the VBM, CBM, and indirect band gap at fixed lattice parameters and neglecting thermal expansion is shown in \cref{fig:temp}.
The zero-point renormalization (ZPR) of the DFT band gap is calculated to be \SI{-0.23}{eV}, with nearly equal contributions from a decrease of the CBM (\SI{-0.12}{eV}) and an increase of the VBM energies (\SI{+0.11}{eV}).
This large correction reduces the DFT-PBE gap from \SI{3.01}{eV} to \SI{2.78}{eV}.

At \SI{300}{K}, the band gap at unit cell volume \Vdft{} is predicted to be reduced by an additional \SI{-0.12}{eV}.
The rate of change of the gap at this temperature is \SI{0.05}{eV}/\SI{100}{K}, and increases only slightly to the linear limit of \SI{0.064}{eV}/\SI{100}{K} at temperatures beyond \SI{500}{K}.

The DFT gap for the experimental room-temperature structure at the enlarged volume \Vexp{} is \SI{3.12}{eV}, an increase of \SI{0.11}{eV} compared to \Vdft{}.
The renormalization calculated at \SI{300}{K} (\SI{-0.44}{eV}) brings it down to \SI{2.68}{eV}.
We observe that the two contributions to the renormalization we compute---the lattice expansion and the zero-temperature contribution from the electron-phonon interaction---are not independent, additive terms.
The EPC shows non-negligible volume dependence, with the renormalization increasing by \SI{26}{\percent} from \SI{-0.35}{eV} at \Vdft{} to \SI{-0.44}{eV} at \Vexp{}.
This can be explained by a narrowing of the electronic bands upon lattice expansion and hence an increase in the electronic DOS.
The increased DOS near and at the band edges leads to more scattering channels on the scale of the phonon energies, and thus an overall larger self-energy.
Altogether, the volume expansion of \Vexp{} leads to two contributions to the renormalization of opposite signs,
  resulting in a band gap at \SI{300}{K} that is only \SI{70}{meV} smaller than the value at \SI{0}{K}.

For a more detailed analysis of the ZPR and temperature dependence, we examine the individual phonon contributions to the renormalization.
Reorganizing \cref*{eq:sigma} we can write
\begin{align}
  \Sigma^\mathrm{ep}_{n\vk}(\omega)
   & =
    \sum_{\nu\vq} \left[ \Sigma^\mathrm{Fan}_{n\vk,\nu\vq}(\omega)
                      + \Sigma^\mathrm{DW}_{n\vk,\nu\vq} \right]
   = \sum_{\nu\vq} \Sigma^\mathrm{ep}_{n\vk,\nu\vq}(\omega)
\end{align}
to obtain the contribution from each phonon.
For this analysis we calculate the self-energy on a $\vq$-grid of \sixeightsix{}, since this phonon decomposition does not hold for our interpolation scheme with two $\vq$-grids.

\begin{figure}
\centering
\includegraphics[]{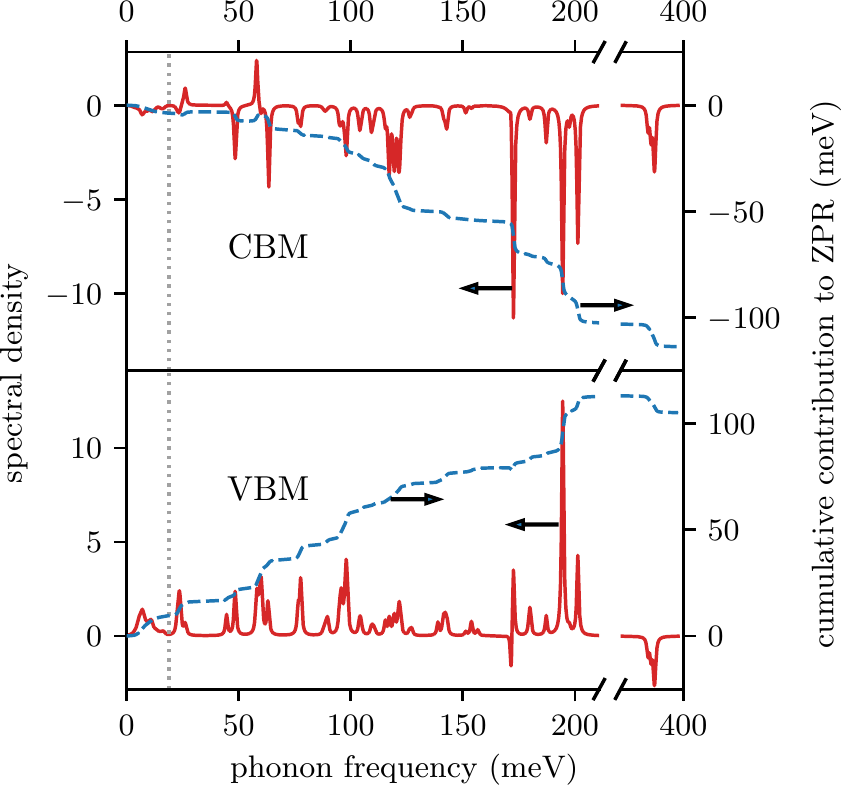}
\caption{\label{fig:phdecomp} Individual contributions of the phonon modes to the renormalization of the CBM and VBM plotted against frequency, with a Lorentzian smearing of \SI{1}{meV} (red solid line, left axis).
The gray dotted line at \SI{19}{meV} indicates the separation of inter- from intramolecular modes.
The blue dashed line (right axis) shows the cumulative integral of the individual contributions.}%
\end{figure}

In \cref{fig:phdecomp} we plot the real part of each $\Sigma^\mathrm{ep}_{n\vk,\nu\vq}(\varepsilon^0_{n\vk})$ at \SI{0}{K}---i.e. each phonon's contribution to the ZPR.
To account for finite sampling of reciprocal space we used a Lorentzian broadening of \SI{1}{meV}.
The intramolecular phonon modes around \SI{190}{meV} are found to have the largest individual contributions, in agreement with previous studies \cite{Vukmirovic2012,Lee2018}.
Overall, however, the contribution as a function of phonon frequency is distributed relatively equally over the frequency range, especially for the VBM, as can be seen from the integral of the spectral density (blue line in \cref{fig:phdecomp}).
The intermolecular modes situated below \SI{19}{meV} (gray dashed line in \cref{fig:phdecomp}) contribute comparatively little to the ZPR.
Only these weakly coupling intermolecular and a few soft intramolecular modes are populated at ambient temperatures, and contribute to the further reduction of the gap at finite temperatures.

A more quantitative description of the fundamental band gap can be achieved by correcting the DFT band gap with many-body perturbation theory within the $GW$ approximation for the self-energy due to electron-electron interaction, then adding the EPC corrections to account for the electron-phonon interaction.
Our previous work shows that the $GW$ method
  increases the indirect DFT band gap of naphthalene
  by about \SI{2.3}{eV}~\cite{Rangel2016},
  thus bringing the band gap of the expanded room-temperature structure to \SI{5.4}{eV}.
Adding the electron-phonon coupling renormalization computed at \SI{300}{K}, we obtain a fundamental gap of \SI{5.0}{eV},
  in excellent agreement with the experimental room-temperature value of \SI{5}{eV}~\cite{Braun1970}.

The electron-electron correlation itself affects the EPC, as reported in prior work, and efforts have been put towards developing methods to capture and quantify this effect.
\cite{Faber2012,Antonius2014,Monserrat2015,Monserrat2016,Li2019}.
Considering the similarity of the magnitudes of electronic bandwidth, phonon, and electron-phonon coupling energies in naphthalene, it is plausible that inclusion of electron-electron correlation has a significant effect on the renormalization; however, we defer this investigation to future work.

\subsection{Electrical mobilities}

We compute the electrical mobilities of the electrons ($\mu^\mathrm{e}$) and the holes ($\mu^\mathrm{h}$) in the self-energy relaxation-time approximation~\cite{Bernardi2014,Giustino2017,Ponce2018} with the expression
\begin{equation} \label{eq:muboltz}
    \mu^\mathrm{e,h}_{\alpha}(T) = \frac{-e}{\rho_\mathrm{e,h} \Omega} \sum_{n} \int \frac{d \vk}{\Omega_\mathrm{BZ}}
    \frac{\partial f(\varepsilon,T)}{\partial \varepsilon}\Big|_{\varepsilon_{n\vk}}
    \vert v_{n\vk,\alpha} \vert^2 \tau_{n\vk}(T),
\end{equation}
where $\alpha$ is the Cartesian direction of the applied electric field and the current, $\rho_\mathrm{e,h}$ is the carrier density of the electrons or the holes, $\Omega$ and $\Omega_\mathrm{BZ}$ are the volumes of the unit cell and the Brillouin zone, $v_{n\vk,\alpha}$ is the velocity of the electronic state $n\vk$ along direction $\alpha$, and the sum over bands is restricted to conduction bands for $\mu^\mathrm{e}$ and valence bands for $\mu^\mathrm{h}$.
The lifetimes $\tau_{n\vk}$ are obtained from the imaginary part of the electron-phonon self-energy
\begin{equation}
\tau_{n\vk}^{-1}(T) =
    \frac{2}{\hbar} 
    \Im\left[
        \Sigma^\mathrm{ep}_{n\vk}(\varepsilon^0_{n\vk},T)
    \right]
.
\end{equation}

To evaluate \cref{eq:muboltz} we use the Wannier90 package~\cite{Mostofi2014} to interpolate our computed electronic eigenvalues and velocities to a $60\times60\times60$ $\vk$-grid.
Calculating the EPC on this fine mesh is prohibitively expensive.
We find, however, that the frequency-dependent self-energy for the bands around the gap is nearly independent of $\vk$ for naphthalene (see the supplemental material for a detailed analysis~\cite{SM}).
We therefore obtain the lifetimes $\tau_{n\vk}$ on the dense $\vk$-grid by interpolating the self-energy $\Sigma^\mathrm{ep}_{n\vk'}$ of a single point $\vk'$ using the approximation
\begin{equation}
 \tau_{n\vk}^{-1}(T)
 \approx \frac{2}{\hbar} \Im[\Sigma^\mathrm{ep}_{n\vk'}(\varepsilon^0_{n\vk},T)].
\end{equation}
To minimize errors associated with this approximation, we choose $\vk'$ to be at A for the hole, and $\Gamma$ for the electron mobility, the locations of the VBM and CBM, respectively.

\begin{table}
\caption{%
Calculated mobilities in comparison with experimental values. We interpolated the experimental results reported in Ref.~\cite{Madelung2000} to \SI{50}{K} and \SI{300}{K}, and compare to calculations using the relaxed (\Vdft{}) and experimental room-temperature volume (\Vexp{}), respectively. Mobility values are given along crystal vectors $a$ and $b$, as well as $c^*$, defined as the vector perpendicular to the $ab$ plane. All values in \si{cm^2/Vs}.
}\label{tab:mobilities_sc}
 \begin{ruledtabular}
\begin{tabular}{lrrr@{\hskip 1.5em}rrr}
{} & \multicolumn{3}{c}{hole} & \multicolumn{3}{c}{electron} \\
{} & $\mu^\mathrm{h}_a$ & $\mu^\mathrm{h}_b$ & $\mu^\mathrm{h}_{c*}$ & $\mu^\mathrm{e}_a$ & $\mu^\mathrm{e}_b$ & $\mu^\mathrm{e}_{c*}$ \\
$T=\SI{50}{K}$ \\
Calc. (\Vdft{}) &        20.03 &        25.73 &            5.84 &        20.45 &         2.74 &            5.02 \\
Exp.            &        65.73 &        68.31 &           35.89 &         7.18 &         3.31 &            0.94 \\[0.5em]
$T=\SI{300}{K}$ \\
Calc. (\Vdft{}) &         3.42 &         4.89 &            0.56 &         2.48 &         0.66 &            0.38 \\
Calc. (\Vexp{}) &         0.96 &         2.24 &            0.20 &         0.61 &         0.29 &            0.19 \\
Exp.            &         0.79 &         1.34 &            0.31 &         0.58 &         0.63 &            0.39 \\
\end{tabular}
 \end{ruledtabular}
\end{table}

The calculated temperature-dependent hole and electron mobilities are shown in \cref{tab:mobilities_sc} for the directions $a$, $b$, and $c^*$ (cf. \cref{fig:naph_structure}).
%
We compare the mobilities at \SI{50}{K} and \SI{300}{K}, using the relaxed (\Vdft{}) and experimental room-temperature volume (\Vexp{}), respectively.
%
Below \SI{50}{K}, the mobilities become dependent on the electric field.
At the same time, the volume between \SI{5}{K} and \SI{50}{K} expands less than \SI{0.5}{\percent},
  and the contribution of thermal lattice expansion to the mobility
  at these temperatures
  is expected to still be negligible.
This allows us to use the relaxed lattice parameters and to extract the contribution of the lattice expansion to the mobility.

At \SI{50}{K}, our calculations generally underestimate the hole mobilities, consistent with prior work \cite{Lee2018}, and overestimate the electron mobilities.
At \SI{300}{K}, the agreement with experiment is reasonably good when using the experimental lattice parameters.
%
This suggests that electronic band transport limited by phonon scattering accounts for much of the electrical mobility.
%
It is also apparent that the lattice expansion plays an important role in obtaining accurate values, as the agreement at \SI{300}{K} greatly improves in most cases when using the room-temperature unit cell with \Vexp{}.
To more accurately predict the power law (or the slope) of the experimental mobilities, calculations need to be repeated using experimental lattice parameters obtained at different temperatures.
This has been shown to lead to good agreement of the power law exponents in prior work~\cite{Lee2018}.
%
%
%
%
%
Possible reasons for any disagreement with experiment include our neglect of polaronic effects and the physics of a hopping transport mechanism.
In particular, at temperatures above \SI{100}{K}, the experimental electron mobilities in the $b$ and $c^*$ direction show a decreased temperature dependence, commonly attributed to the transition to hopping transport~\cite{Schein1978,Schein1979,Warta1985,Ortmann2010} (see also the supplemental material~\cite{SM}).
Nonetheless, our work can be considered an important baseline for comparing with experiments and future work incorporating polaronic effects.
%

To gain insight into the mobilities, we decompose them into energy-resolved contributions
by approximating \cref{eq:muboltz} in the following way
\begin{equation} \label{eq:muapprox}
    \mu^\mathrm{e,h}_{\alpha} \approx \frac{-e}{\rho_\mathrm{e,h}}
    \int \intd \varepsilon\, D(\varepsilon) f'(\varepsilon) v_{\alpha}^2(\varepsilon) \tau(\varepsilon),
\end{equation}
where $D(\varepsilon$) is the density of states (DOS), $f'(\varepsilon)$ is the derivative of the Fermi-Dirac distribution with respect to energy, and where we define the average squared velocity function
\begin{equation} \label{eq:v2approx}
    v_{\alpha}^2(\varepsilon) = \frac{1}{D(\varepsilon)}\sum_n \int \frac{d\vk}{\Omega_\mathrm{BZ}} ( v_{n\vk,\alpha} )^2 \delta(\varepsilon - \varepsilon_{n\vk}),
\end{equation}
and the average lifetime function
\begin{equation} \label{eq:tauapprox}
    \tau(\varepsilon) = \frac{1}{D(\varepsilon)}\sum_n \int \frac{d\vk}{\Omega_\mathrm{BZ}} \tau_{n\vk} \delta(\varepsilon - \varepsilon_{n\vk})
.
\end{equation}
The bounds of the integral in \cref{eq:muapprox} go from $-\infty$ to the Fermi energy $\varepsilon_\mathrm{F}$ for holes, and from $\varepsilon_\mathrm{F}$ to $+\infty$ for electrons, and we add a small Gaussian smearing of \SI{5}{meV} to evaluate the Dirac delta functions in \cref{eq:v2approx,eq:tauapprox}.

\begin{figure*}
  \includegraphics[]{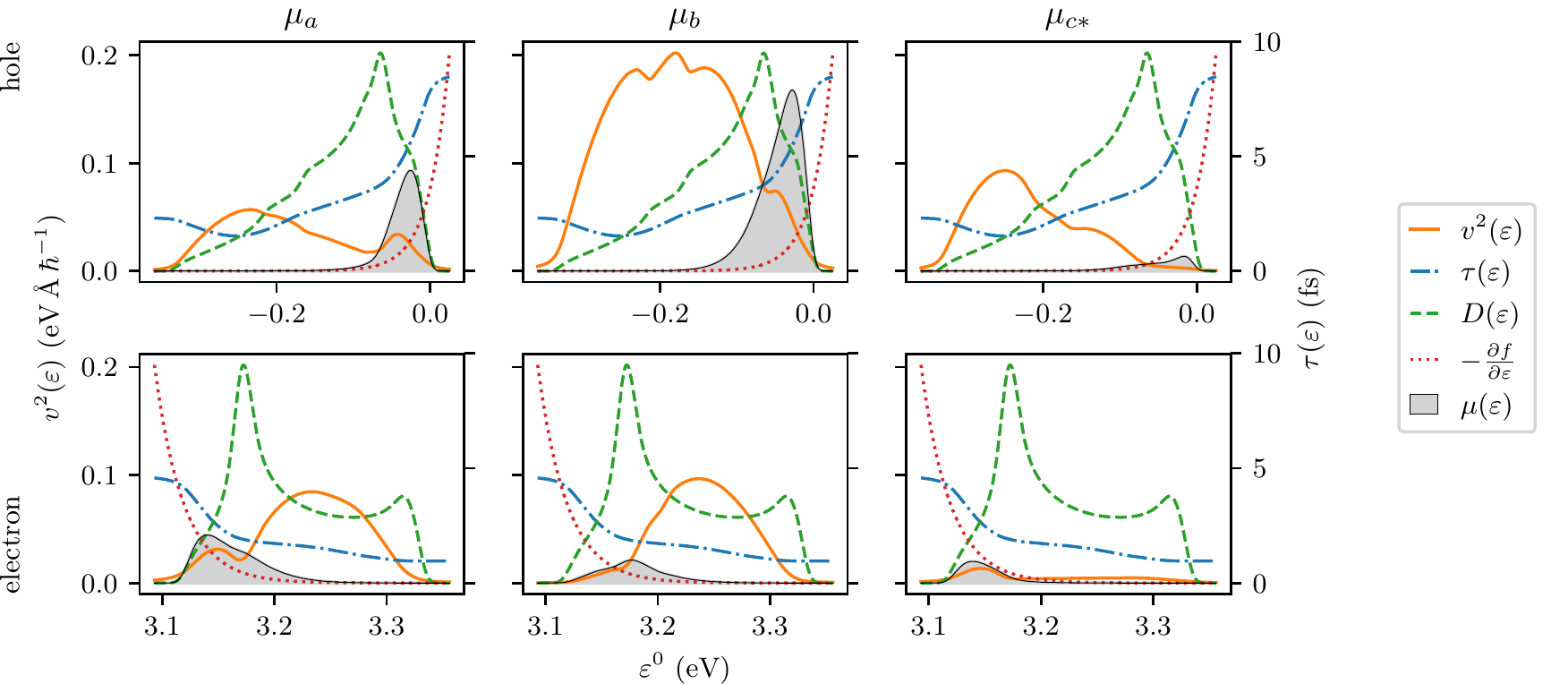}
  \caption{%
  The energy-resolved decomposition of the mobility according to \cref{eq:muapprox} of holes (top) and electrons (bottom) at \SI{300}{K} and the experimental room-temperature structure with \Vexp{}. The velocity (orange solid) and the lifetime (blue dash-dot) are associated with the left and right $y$-axes respectively. The density of states $D(\varepsilon)$ (green dashed), the derivative of the Fermi-Dirac distribution (red dotted), and the mobility integrand (gray filled) are in arbitrary units, but share the same scale across all plots.
}\label{fig:mobilities_decomposition}
\end{figure*}

\Cref{eq:muapprox} approximates the energy-resolved contributions to the mobilities as the product of four functions of energy.
We plot these quantities for \Vexp{} in \cref{fig:mobilities_decomposition}.
At \SI{300}{K}, the contributions to the mobilities extend up to about \SI{0.1}{eV} above or below the band edges.
Within this region, the DOS, velocity, and lifetime are generally not monotonic functions of energy, but show distinct features.
This highlights the need for our detailed calculations; in contrast, for example, approximations of the mobility that only use the effective mass of the band extrema, or constant effective lifetimes will be inadequate.
This is especially true for $\mu^\mathrm{e}_b$, where the main contribution to the mobility
  is situated near the peak of the DOS, almost \SI{0.1}{eV} within the conduction band.
Using this analysis, we can also explain why the electron mobilities are generally lower than the hole mobilities.
Comparing the individual quantities, we see that the velocities of electrons along the $a$ and $c^*$ directions are actually larger than those of the holes.
However, the lower electron lifetimes compared to the hole lifetimes, especially near the band edge, more than compensate for the higher velocities.
In general, this analysis shows the critical role the individual contributions of \cref{eq:muboltz} play in quantitatively determining the mobility.

While the expression in \cref{eq:muapprox} is of great practicality for computing the mobilities and visualizing the energy-resolved lifetimes and velocities, it also turns out to be an excellent approximation.
The maximum relative error compared to \cref{eq:muboltz} is below \SI{10}{\percent}, and the mean absolute relative error is below \SI{5}{\percent}.
Mobilities calculated with this approximation deviate less than \SI{3.3}{\percent} (see the supplemental material \cite{SM}).
In addition to being independent of $\vk$, the frequency-dependent self-energies of the two highest (lowest) valence (conduction) bands are almost identical.
This is because the wave functions, and hence the electron-phonon matrix elements, of Davydov pairs are so similar for naphthalene (see \cite{SM}).
Within this $\vk$- and $n$-independent approximation, the electron and hole lifetimes are only a function of energy, and the expressions in \cref{eq:muboltz} and \cref{eq:muapprox} become equivalent.

\subsection{Self-consistent electron-phonon self-energy}

\begin{figure}
  \includegraphics[]{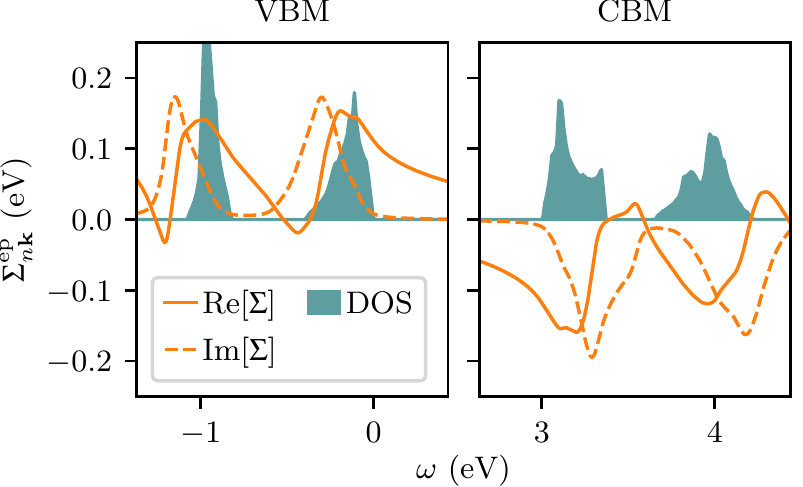}
  \caption{%
    The real (solid) and imaginary part (dashed) of the electron-phonon self-energy of naphthalene, evaluated for the VBM at A (left) and CBM at $\Gamma$ (right). The features of the self-energy correlate with the electronic DOS (filled).}
  \label{fig:self_energy}
\end{figure}

\Cref{fig:self_energy} shows the frequency-dependent electron-phonon self-energy of the valence and conduction band extrema alongside the electronic DOS.
We see a clear correlation.
%
%
%
%
This is mainly due to the fact that the electron-phonon coupling matrix elements are relatively independent of $\vk$ and $n$ within a Davydov pair.
The imaginary part of \cref{eq:Fan} then  becomes proportional to the joint electronic and vibrational density of states,
  weighted by the coupling strength of each phonon.
In agreement with previous studies \cite{Vukmirovic2012,Lee2018},
we find that intramolecular modes around \SI{0.19}{eV} have the strongest coupling
(\cref{fig:phdecomp}).
Correspondingly, the peaks of the imaginary part of the SE are shifted by about \SI{0.19}{eV} compared to the peaks of the DOS.

We also note from \cref{fig:self_energy} that the real part of the electron-phonon self-energy varies rapidly between 0 and \SI{0.15}{eV}
  over the frequency range corresponding to the band width, which is on the order of \SI{0.4}{eV}.
The renormalization of the bands will therefore significantly alter the shape and width of the DOS,
  upon which the self-energy depends.
The magnitude of the self-energy corrections
  suggests that we should compute the self-energy self-consistently,
  by updating the electronic energies in \cref{eq:Fan} with the renormalized values.

Accordingly, we use an eigenvalue--self-consistent (evSC) cycle for the self-energy,
  whose iterative steps can be summarized as
\begin{align}
\varepsilon^1_{n\vk} 
    &= \varepsilon^0_{n\vk} + \Re 
    \left[ 
    \Sigma^\mathrm{ep}_{n\vk}(\varepsilon^0_{n\vk}, \varepsilon^0_{m\vk+\vq})
    \right]
    \notag
    \\
\varepsilon^2_{n\vk}
    &= \varepsilon^0_{n\vk} + \Re 
    \big[
    \Sigma^\mathrm{ep}_{n\vk}(\varepsilon^1_{n\vk}, \varepsilon^1_{m\vk+\vq})
    \big]
    \notag
    \\
    \ldots \notag
    \\
\varepsilon^i_{n\vk}
    &= \varepsilon^0_{n\vk} + \Re 
    \big[
    \Sigma^\mathrm{ep}_{n\vk}(\varepsilon^{i-1}_{n\vk}, \varepsilon^{i-1}_{m\vk+\vq})
    \big],
\end{align}
where $\Sigma^\mathrm{ep}_{n\vk}(\varepsilon^{i-1}_{n\vk}, \varepsilon^{i-1}_{m\vk+\vq})$ indicates the use of renormalized eigenvalues in the self-energy.
We use the $\vk$-independence approximation to efficiently calculate the renormalized states $m\vk\!+\!\vq$ as
\begin{equation}
    \varepsilon^i_{m\vk+\vq}
    \approx \varepsilon^0_{m\vk+\vq} + 
        \Re \left[ \Sigma^\mathrm{ep}_{n\vk}(\varepsilon^{i-1}_{m\vk+\vq}) \right].
\end{equation}
This procedure converges the renormalized energies rapidly to within \SI{2}{meV}
  for the bands around the gap (see the supplemental material \cite{SM}).

Our method effectively includes all high-order non-crossing electron-phonon coupling diagrams
  in the self-energy.
It does not, however, allow for multi-phonon satellite bands to form in the spectral function, as, for example, the cumulant expansion would~\cite{Nery2018}.
A similar level of theory to evSC was previously achieved using a time propagation
  of the Green's function~\cite{Dunn1975}.

While the self-consistent calculation of the electron-phonon coupling self-energy offers a clear description of the quasiparticle temperature renormalization and lifetimes, one generally aims to compute these quantities from a one-shot calculation of the self-energy for practical reasons.
Two different procedures are often used.
In the on-the-mass-shell approximation~\cite{Cannuccia2012}, which we have used so far,
the renormalized energies are computed according to \cref{eq:renorm}.
%
A more rigorous approach, in theory, is to evaluate the self-energy at the quasiparticle energy, corresponding to the peak of the spectral function, that is,
\begin{equation} \label{eq:Dyson}
\varepsilon_{n\vk}(T) = \varepsilon^0_{n\vk} + \Re[ \Sigma^\mathrm{ep}_{n\vk}(\varepsilon_{n\vk},T)].
\end{equation}
In \cref{tab:oneshotsc}, we compare the two one-shot procedures against the self-consistent scheme.
For the VBM and the CBM, the on-the-mass-shell approximation appears to better reproduce the self-consistent scheme, both for the real and imaginary parts of the self-energy.
The quasiparticle solution vastly overestimates the lifetimes of the band extrema 
(see the supplemental material \cite{SM}).
For the real part of the self-energy, such a result agrees with the one found for the Fr\"ohlich model, compared with diagrammatic Monte Carlo results~\cite{Mishchenko2000,Nery2018}.

\begin{table}
 \begin{ruledtabular}
\caption{%
Comparison of the one-shot self-energy computed in the on-the-mass-shell approximation ($\Sigma(\varepsilon^0)$),
the one-shot self-energy evaluated at the quasiparticle solution
($\Sigma(\varepsilon)$),
and the eigenvalue--self-consistent self-energy (evSC).
Renormalizations $\Delta \varepsilon$ are in \si{eV}, lifetimes $\tau$ in \si{fs}.
}\label{tab:oneshotsc}
\begin{tabular}{lrrr}
{} &  {$\Sigma(\varepsilon^0)$} &  {$\Sigma(\varepsilon)$} &  {evSC} \\[0.5em]
$\Delta \varepsilon_\mathrm{VBM}$ (\SI{0}{K}) &               0.11 &             0.09 &  0.12 \\
$\Delta \varepsilon_\mathrm{CBM}$ (\SI{0}{K}) &              -0.12 &            -0.09 & -0.12 \\
$\tau_\mathrm{VBM}$ (\SI{300}{K})             &               8.70 &            38.47 &  7.91 \\
$\tau_\mathrm{CBM}$ (\SI{300}{K})             &               4.73 &            21.16 &  6.42 \\
\end{tabular}
\end{ruledtabular}
\end{table}

Next, we examine the effect of the evSC approach through the spectral function,
given by the imaginary part of the Green's function:
\begin{equation} \label{eq:sf}
  A_{n\vk}(\omega) = 
    \frac{1}{\pi} 
    \frac{\big| \Im\left[\Sigma^\mathrm{ep}_{n\vk}(\omega)\right]\big|}
         {\big[ \omega - \varepsilon^0_{n\vk} 
                - \Re\left[\Sigma^\mathrm{ep}_{n\vk}(\omega)\right]\big]^2
          + \Im\left[\Sigma^\mathrm{ep}_{n\vk}(\omega)\right]^2
         }.
\end{equation}
It describes the probability of finding an electron in state $n\vk$ at energy $\omega$.
The quasiparticle (QP) peaks of the spectral function appear at $\omega = \varepsilon^0 - \Re[\Sigma^\mathrm{ep}_{n\vk}(\omega)]$, which corresponds to the solution of \cref{eq:Dyson}.
The spectral function allows us to compare both the renormalization (position of the QP peak) and the broadening (width and height of the QP peak) simultaneously.

\begin{figure}
  \includegraphics[]
  {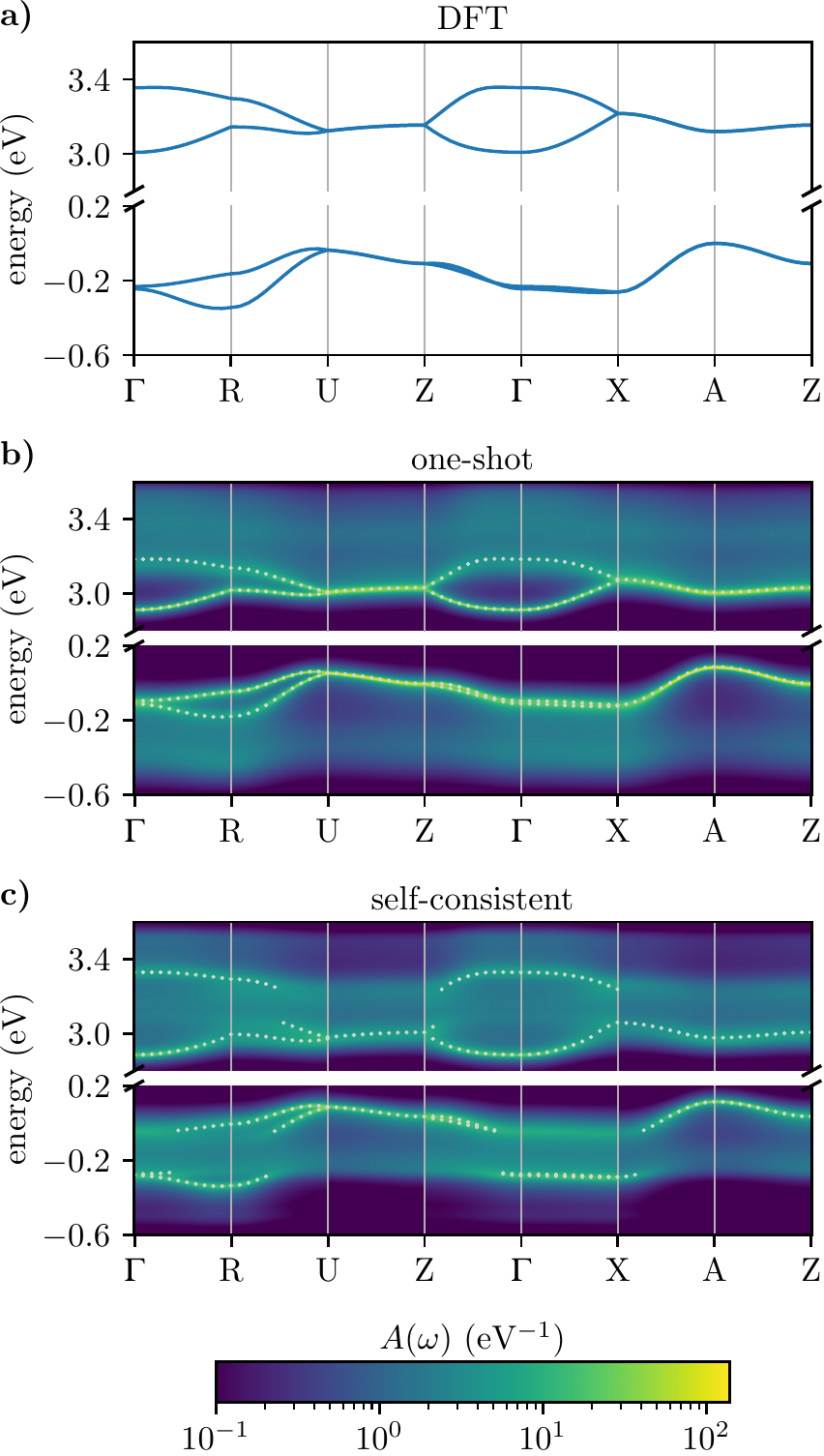}
  \caption{%
  (a) The DFT-PBE-D3 band structure of naphthalene of the two highest valence, and two lowest conduction bands.
  (b) and (c) The spectral function of the full band structure calculated using the (b) one-shot and (c) self-consistent method.
  To highlight the renormalized band structure, the highest peak for each state $n\vk$, i.e., the solution to \cref{eq:Dyson} with the smallest imaginary part, is marked with a dot.
  While the one-shot spectral function displays a continuous quasiparticle band-structure, the self-consistent result shows discontinuities.
  }\label{fig:sf_full_bs}
\end{figure}

\Cref{fig:sf_full_bs} shows both the one-shot and evSC spectral function, where we use the $\vk$-independence approximation to interpolate $A_{n\vk}(\omega)$ across the Brillouin zone.
We chose the self-energy at $\Gamma$ as a starting point for the interpolation, and we checked that the choice of starting point does not alter the results significantly.

The QP bands of the evSC spectral function show a discontinuity at energies around \SI{0.2}{eV}
  below the VBM and above the CBM, 
  due to the spectral weight being transferred from the main quasiparticle peak to the satellite band.
In contrast, the bands of the one-shot calculation are continuous,
  and the distinction between the main quasiparticle peak and the satellite
  remains clear in most cases.
This band discontinuity (or splitting) happens when the real part of the self-energy has a unitless slope $\gtrsim 1$.
In this case, the Dyson equation~\eqref{eq:Dyson} may admit more than one solution in certain regions of the Brillouin zone.
Such a high slope in the self-energy is seen near the poles, located one phonon frequency away from the peaks of the DOS, as seen in \cref{fig:self_energy} (the strongest coupling modes are \SI{\sim0.19}{eV}).
A similar splitting has also been observed theoretically and experimentally in pentacene and rubrene crystals~\cite{Ciuchi2011,Ciuchi2012,Bussolotti2017}
as well as non-organic systems~\cite{Eiguren2008,Eiguren2009}.

Finally, we evaluate the mobilities from the evSC self-energy at \SI{300}{K} using \Vexp{} lattice parameters,
  taking into account the renormalized electronic eigenvalues and velocities.
The results are listed in \cref{tab:mobilities} in comparison with the values for the one-shot calculation and experiment.
The evSC approach lowers the hole mobilities, bringing $\mu_a$ and $\mu_b$ to even better agreement with experiment.
In contrast, evSC electron mobilities increase slightly compared to the one-shot calculation.
By looking at the decomposition of the mobility via \cref{eq:muapprox},
  we can attribute the decrease of the hole mobility to lower lifetimes,
  and the increase of the electron mobilities to higher lifetimes and velocities
  (see the supplemental material \cite{SM} for the decomposition).

\begin{table}
 \caption{%
Mobilities calculated at \SI{300}{K} with experimental lattice parameters (\Vexp{}), using the one-shot and self-consistent (evSC) method, in comparison with experimental values.
 All values in \si{cm^2/Vs}.
 }\label{tab:mobilities}
\begin{ruledtabular}
\begin{tabular}{l*{3}{r}@{\hskip 3em}*{3}{r}}
{} & \multicolumn{3}{c}{hole} & \multicolumn{3}{c}{electron} \\
{} & {$\mu_a$} & {$\mu_b$} & {$\mu_{c^*}$} & {$\mu_a$} & {$\mu_b$} & {$\mu_{c^*}$} \\[0.3em]
one-shot &    1.20 &    2.73 & 0.24 &   0.67 & 0.31 & 0.21 \\
evSC       &    0.90 &    2.19 & 0.18 &   1.18 & 0.59 & 0.31 \\
Exp.     &    0.79 &    1.34 & 0.31 &   0.58 & 0.63 & 0.39 \\
\end{tabular}
\end{ruledtabular}
\end{table}


\section{Conclusion}

In summary, we used comprehensive \textit{ab initio} calculations based on DFT to study the effect of electron-phonon interactions on the electronic structure
  of naphthalene crystals, as well as its electrical mobility.
%
%
Both the temperature-dependent renormalization of the gap, and the hole and electron mobilities
  are in good agreement with experimental values, if the lattice expansion is taken into account.
Because of the limited dependence of the self-energy on $\vk$ and $n$ of the two occupied and unoccupied band-edge bands,
  we can visualize the contributions to the mobility at each band energy
  in terms of the density of states, average scattering time, and average velocity squared.
This facilitates a useful energy-resolved analysis of the mobility,
  and provides an efficient way to model charge carrier transport in organic systems.

Furthermore, we indirectly and approximately investigated the effect of higher-order electron-phonon coupling terms
  by calculating the self-energy self-consistently.
The band gap renormalization and mobilities show only moderate differences between the one-shot and self-consistent calculations,
  as long as the on-the-mass-shell approximation is used.
Both of these properties depend mainly on the electronic states close to the band gap, which are only weakly affected by the evSC treatment.
However, the electronic states further away from the band edges are strongly affected
  by the self-consistent treatment of the self-energy.
The spectral function reveals a band splitting and band widening 
  comparable to what has been observed experimentally in other molecular crystals.
%
%

Most of the qualitative results discussed in this work result directly
  from the weak interactions between constituent monomers, a common feature of molecular crystals.
This includes the $\vk$-independence of the self-energy,
  and the band widths being on the same order of magnitude as the phonon frequencies.
%
The methods and conclusions presented here likely apply to several other molecular crystals,
  and provide an efficient approach for the \textit{ab initio} calculation of the electron-phonon self-energy
  and electrical mobility.


\begin{acknowledgments}
This work was supported by the Theory FWP at the Lawrence Berkeley National Laboratory, which is funded by the US Department of Energy (DOE), Office of Science, Basic Energy Sciences, Materials Sciences and Engineering Division under Contract DE-AC0205CH11231,
and by the Fonds de la Recherche Scientifique (F.R.S.-FNRS Belgium) through the PdR Grants No. T.0238.13 - AIXPHO (X.G., M.G.),
    and No. T.0103.19 - ALPS (X.G., M.G.).
Computational resources were provided by the National Energy Research Scientific Computing Center, which is supported by the Office of Science.
CD acknowledges support by the Deutsche Forschungsgemeinschaft (DFG) - Projektnummer 182087777 - SFB 951.
\end{acknowledgments}

%
%
%
%

\bibliography{main}

\end{document}



\title{Band gap renormalization, carrier mobilities, and the electron-phonon self-energy in crystalline naphthalene}


\author{Florian Brown-Altvater}
\email[]{altvater@berkeley.edu}
\affiliation{Department of Chemistry, University of California, Berkeley, 
  California, USA}
\affiliation{Molecular Foundry, Lawrence Berkeley National Laboratory,
  Berkeley, California, USA}

\author{Gabriel Antonius}
\affiliation{D\'epartement de Chimie, Biochimie et Physique, Institut de recherche sur l'hydrog\`ene, Universit\'e du Qu\'ebec \`a Trois-Rivi\`eres, C.P. 500, Trois-Rivi\`eres,  Canada}
\affiliation{Department of Physics, University of California, Berkeley, 
  California, USA}

\author{Tonatiuh Rangel}
\affiliation{Molecular Foundry, Lawrence Berkeley National Laboratory, 
  Berkeley, California, USA}
\affiliation{Department of Physics, University of California, Berkeley, 
  California, USA}

\author{Matteo Giantomassi}
\affiliation{Universit\'e catholique de Louvain, Louvain-la-Neuve, Belgium}

\author{Claudia Draxl}
\affiliation{Humboldt Universit\"at Berlin, Germany}

\author{Xavier Gonze}
\affiliation{Universit\'e catholique de Louvain,
Louvain-la-neuve, Belgium}
\affiliation{Skolkovo Institute of Science and Technology, Moscow, Russia}

\author{Steven G. Louie}
\affiliation{Department of Physics, University of California, Berkeley, 
  California, USA}
\affiliation{Materials Sciences Division, Lawrence Berkeley National Laboratory, 
  Berkeley, California, USA}

\author{Jeffrey B. Neaton}
\email[]{jbneaton@berkeley.edu}
\affiliation{Molecular Foundry, Lawrence Berkeley National Laboratory, 
  Berkeley, California, USA}
\affiliation{Department of Physics, University of California, Berkeley, 
  California, USA}
\affiliation{Kavli Energy NanoSciences Institute at Berkeley, California, USA}

\date{\today}

\maketitle

\section{Lattice parameters}

\begin{table}[H]
 \caption{\label{SI:tab:lattices}%
 Lattice parameters used in this work. Experimental crystal structures are available at the Cambridge Structural Database \cite{Thomas2010,*csd_url}. The identifiers for the structures measured at \SI{5}{K} and \SI{295}{K} are NAPHTA31 and NAPHTA36, respectively, and published in association with \cite{Capelli2006}. (Lengths in \si{\angstrom}, angles in degrees, volumes $\Omega$ in \si{\angstrom^3}.)}
\begin{ruledtabular}
\begin{tabular}{lrrrrr}
{} &   $a$ &   $b$ &   $c$ & $\beta$ & $\Omega$ \\
PBE-D3 (\Vdft{})           & 8.077 & 5.900 & 8.620 &  124.35 & 339.14 \\
Exp. \SI{{5}}{{K}}                       & 8.080 & 5.933 & 8.632 &  124.65 & 340.41 \\
Exp. \SI{295}{K} (\Vexp{}) & 8.256 & 5.983 & 8.677 &  122.73 & 360.56 \\
\end{tabular}
\end{ruledtabular}
\end{table}

\section{Phonon frequencies}

\begin{figure}[H]
   \centering
   \includegraphics[]{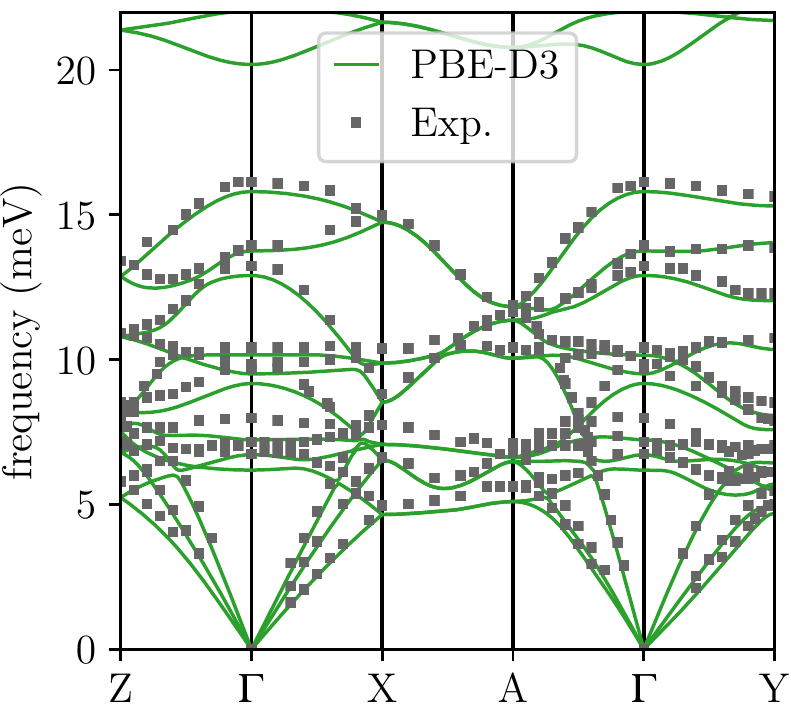}
    \caption{\label{SI:fig:phonons_h8}%
    Phonon band structure of intermolecular bands of perdeuterated naphthalene calculated with PBE-D3 at \Vdft. Experimental neutron scattering frequencies are taken from Ref.~\cite{Natkaniec1980}.
    }
\end{figure}

\begin{figure}[H]
    \centering
  \includegraphics[]{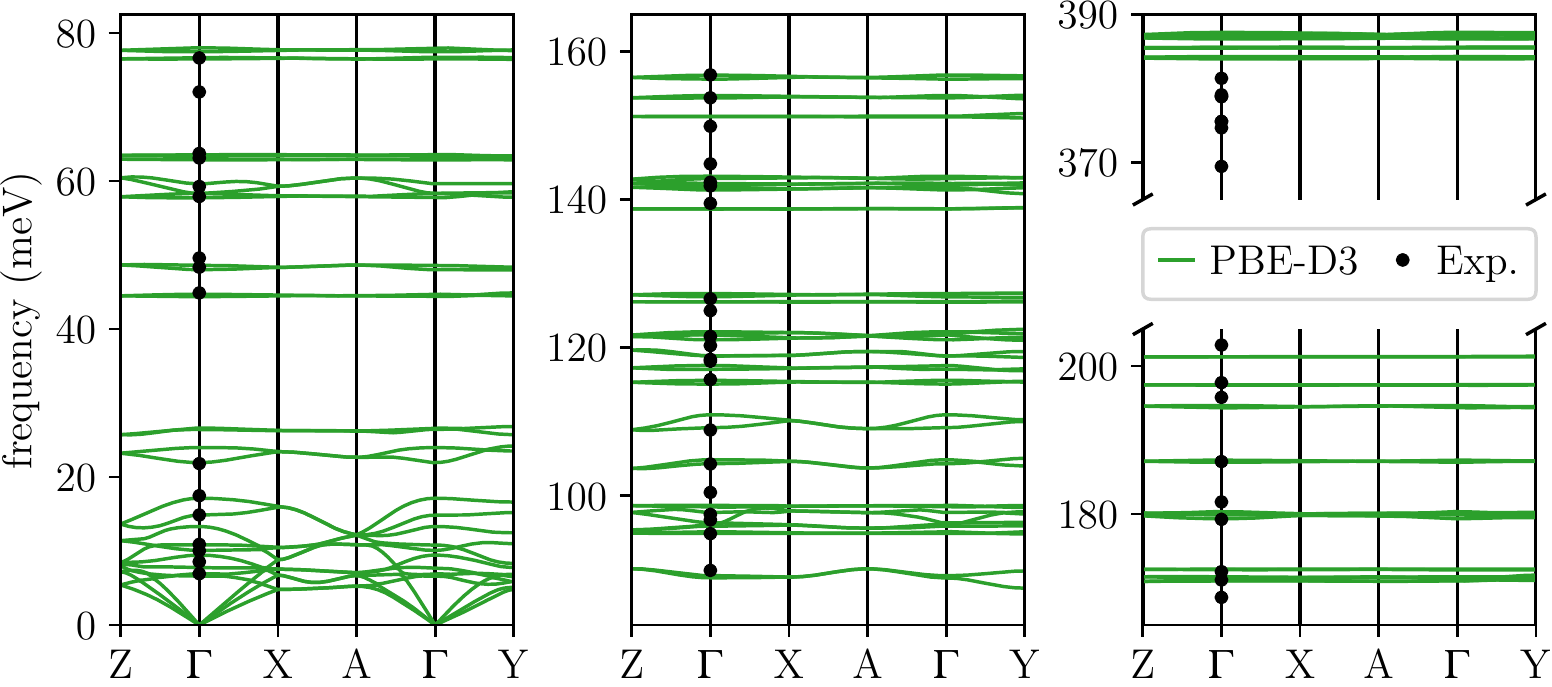}
    \caption{\label{SI:fig:phonons_d8}%
    Phonon band structure of naphthalene calculated with PBE-D3 at \Vdft. Experimental Raman and IR frequencies are taken from Ref.~\cite{Suzuki1968}.
    The discrepancy of the highest phonon frequencies above \SI{380}{eV} coming from C-H stretch modes likely is caused by anharmonic effects, which are ignored by the perturbative approach used here.
    Because these phonon modes contribute only very little to the electron-phonon coupling, the discrepancy of about \SI{5}{\%} of these frequencies will not affect our results in any significant way.
    }
\end{figure}

\begin{figure}[H]
    \centering
  \includegraphics[]{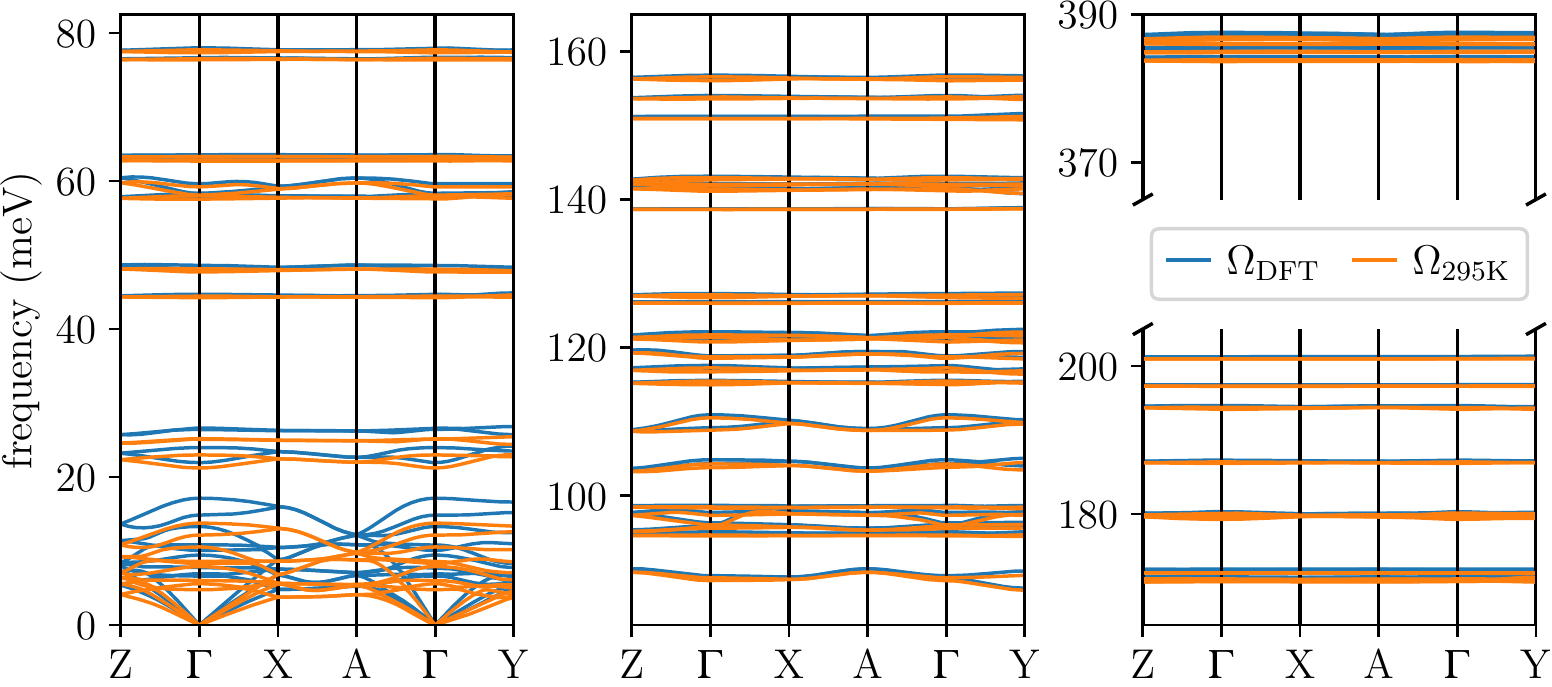}
    \caption{\label{SI:fig:phonons_dos}%
    Comparison of phonon band structures calculated with lattice parameters relaxed with PBE-D3 (\Vdft) and with fixed experimental room temperature lattice parameters (\Vexp). Both calculations are done with PBE-D3. The lattice parameters mainly affect the soft intermolecular modes below \SI{20}{meV}.
    }
\end{figure}


\section{Convergence of self-energy}

\begin{figure}[H]
   \centering
   \includegraphics[]{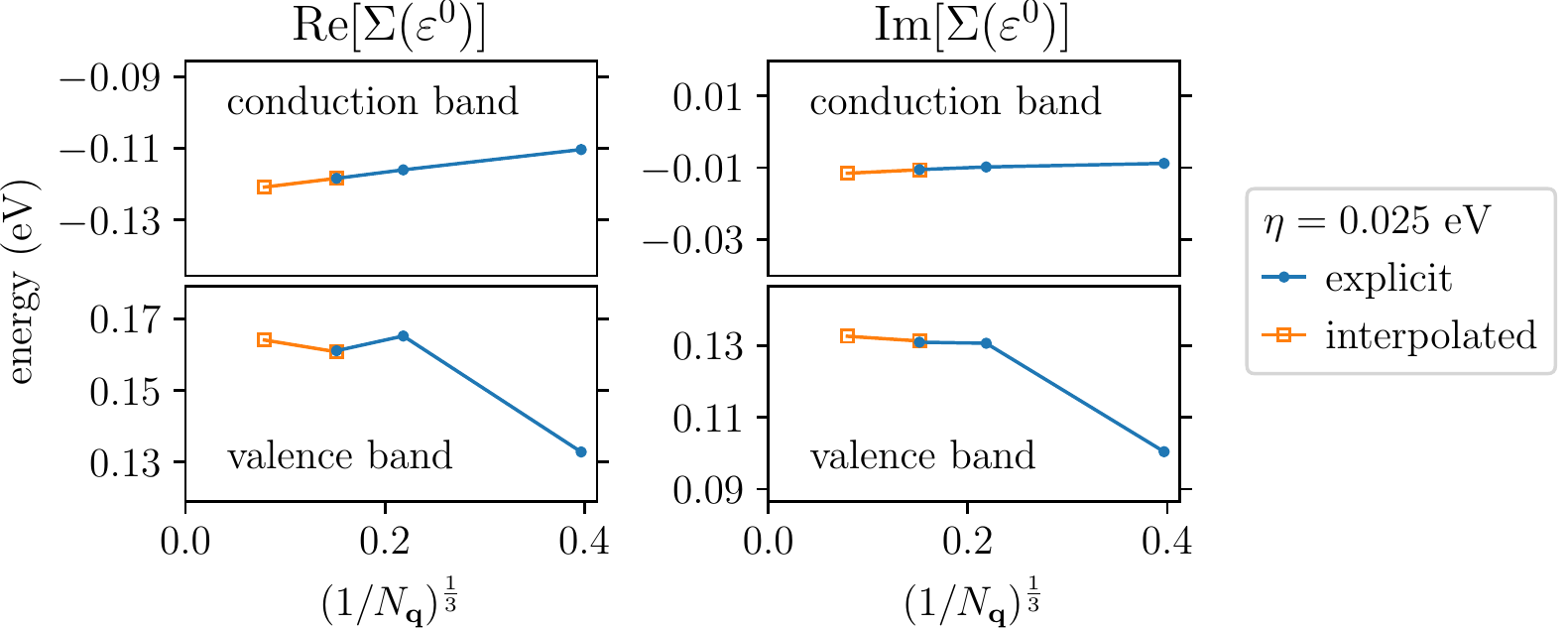}
    \caption{\label{SI:fig:conv_nq}%
        Convergence w.r.t. $\vec{q}$-grid spacing of the electron-phonon self-energy of the highest valence and lowest conduction band at $\Gamma$. The four $\vq$ grids used from right to left are (\twofourtwo), (\foursixfour), (\sixeightsix), and (\twelvefourteentwelve). Interpolated grids use square markers.
        Notable is the much smoother convergence of the conduction band, which is the absolute minimum at $\Gamma$.
        The valence band at $\Gamma$ is \SI{0.23}{eV} below the valence band maximum.
    }
\end{figure}

\begin{figure}[H] 
  \centering
    \includegraphics[]{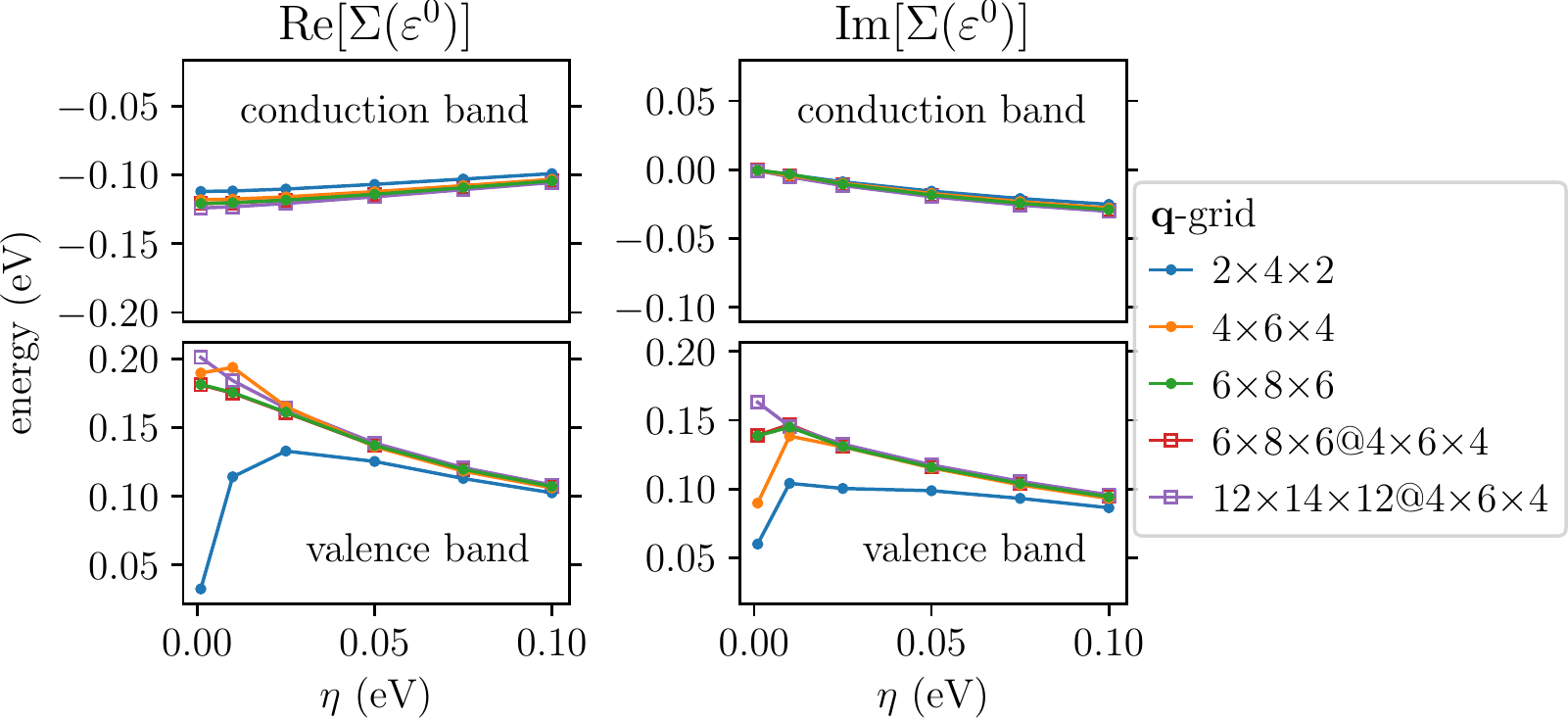}
    \caption{\label{SI:fig:conv_eta}%
        Convergence w.r.t. the smearing value $\eta$ of the electron-phonon self-energy of the highest valence and lowest conduction band at $\Gamma$. Interpolated grids use square markers and are labeled as ``fine@coarse'' to indicate the $\vq$-grids used.
        Notable is the much smoother convergence of the conduction band, which is the absolute minimum at $\Gamma$.
        The valence band at $\Gamma$ is \SI{0.23}{eV} below the valence band maximum.
    }
\end{figure}

\section{\textbf{k}-independence of self-energy}
\label{SI:sec:kindependence}

\begin{figure}[H]
  \centering
  \includegraphics[width=.99\textwidth]{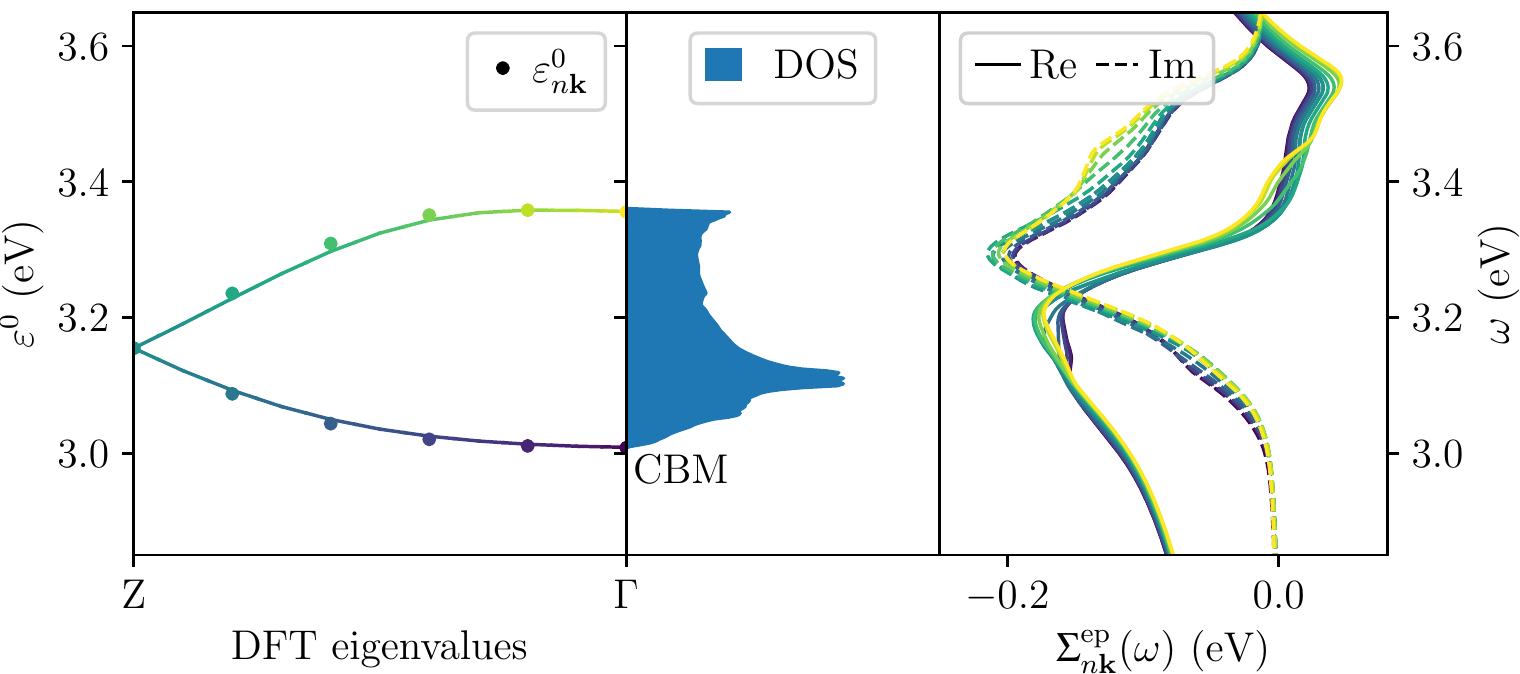}
  \caption{\label{SI:fig:se_LUMO}%
    (Left) Electronic band structure of the two lowest conduction bands along $\Gamma\rightarrow\mathrm{Z}$ in the BZ.
    Each circle indicates an electronic state $n\vk$ for which we explicitly calculated the self-energy $\Upsigma^\mathrm{ep}_{n\vk}$.
    (Middle) The density of states in the middle is plotted to highlight the correlation to the self-energy as discussed in the main text.
    (Right) The real and imaginary part of the electron-phonon self-energy of naphthalene for the states circled in the band structure plot on the left. The color of each self-energy matches the corresponding circle.
    }
\end{figure}

Here we test the $\vk$-independence approximation of the electron-phonon self-energy used in the main text.
In \cref{SI:fig:se_LUMO} we compare the frequency-dependent self-energy along the $\Gamma\rightarrow\mathrm{Z}$ path, the direction of the largest dispersion for the conduction bands.
Even along this relatively dispersive $\vk$-path, the real and imaginary parts of the self-energy show very little variation, validating the $\vk$-independence approximation.
Additionally we can also see that the frequency-dependent self-energy varies very little across the two bands.
The combined $\vk$- and $n$-independence leads to the excellent agreement between the mobilities calculated with \cref*{eq:muboltz} and \cref*{eq:muapprox}.

As laid out in the main text, the two highest valence (lowest conduction) band wavefunctions of the naphthalene crystal, which has two molecules per unit cell, resemble linear combinations of the HOMO (LUMO) of a single naphthalene molecule in the gas phase.
This ``duality'' or pairing of wavefunctions is also called Davydov splitting, or Davydov pairs \cite{Davydov1964,Sheka1975}.
The electronic states of these Davydov pairs around the band gap (for example HOMO/HOMO-1, HOMO-2/HOMO-3, or LUMO/LUMO+1) interact and mix only very weakly with their neighboring Davydov pairs, since they are energetically far enough apart.
Therefore, the resulting wave functions retain their gas-phase-like character throughout the Brillouin zone.
In summary, the two wave functions $n$ and $n+1$ of a Davydov pair have a similar molecular orbital character, and are largely independent of $\vk$.

Since the electron-phonon matrix elements are a measure of the overlap of the electronic wave function with the first derivative of the phonon potential, we expect the same $n$/$n+1$ and $\vk$-independence for the matrix elements.
Additionally, the separation between Davydov pairs near the band gap is on the order of \SIrange{0.3}{0.4}{eV}, and thus at the upper end of our phonon spectrum.
Hence, significant contributions to the self-energy, i.e., terms with small denominators in \cref*{eq:Fan}, will only come from scattering within these Davydov pairs.
These two factors ($\vk$- and $n$-independence of the matrix elements; and well separated bands) lead to the $\vk$- and $n$-independence of the frequency-dependent self-energy.
Since both these properties are very typical for organic crystals, we expect to see this approximation to hold for other systems as well.

\vfill

\section{Electrical mobility}

\begin{figure}[H]
 \includegraphics[]{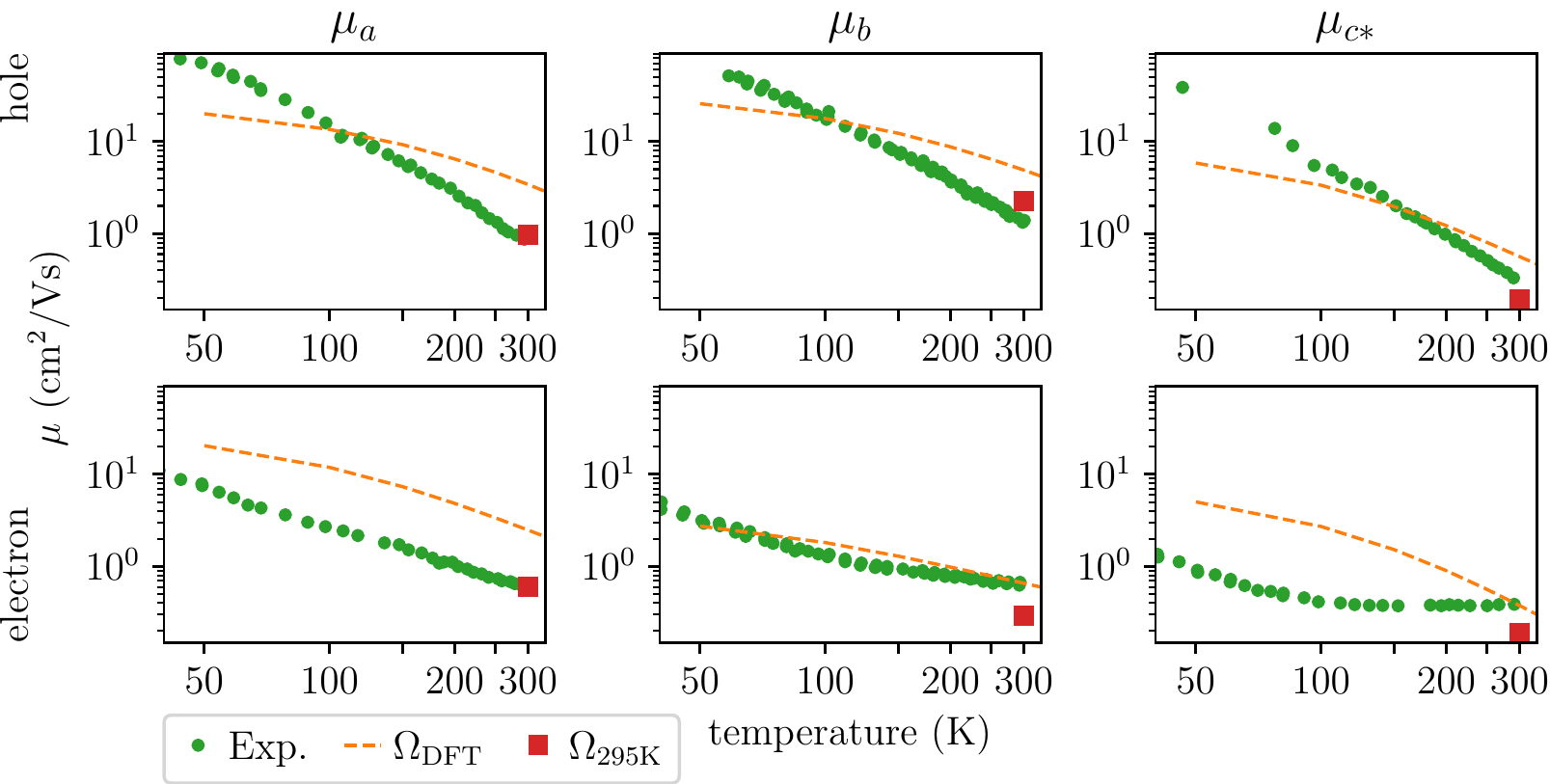}
 \caption{%
    Calculated temperature-dependent hole (top) and electron (bottom) mobilities in comparison with experiment (green dots)~\cite{Madelung2000}. The calculated mobilities at \SI{300}{K} using experimental room temperature (\Vexp{}) lattice parameters (red squares) agree very well with the experimental values.
 For reference, we also show the temperature dependence of the mobilities for the relaxed structure (\Vdft{}, orange dashed).
 }\label{fig:tdep_mobility}
\end{figure}

\begin{table}[H]
\begin{ruledtabular}
\caption{\label{SI:tab:muapprox}%
Mobility values calculated with \Vexp{} at \SI{300}{K}, comparing the values $\mu_\mathrm{Boltz}$ obtained with the self-energy relaxation time approximation \cref*{eq:muboltz}, with $\mu(\varepsilon)$ from \cref*{eq:muapprox}, where we express the mobility as product of four independent functions of energy $
    \mu^\mathrm{e,h}_{\alpha} \propto
    \int \intd \varepsilon\, D(\varepsilon) f'(\varepsilon) v_{\alpha}^2(\varepsilon) \tau(\varepsilon)$.
Mobility values are given in \si{cm^2/Vs}.
}
\begin{tabular}{lrrr@{\hskip 3em}rrr}
{} & \multicolumn{3}{c}{hole} & \multicolumn{3}{c}{electron} \\
{} & $\mu^\mathrm{h}_a$ & $\mu^\mathrm{h}_b$ & $\mu^\mathrm{h}_{c*}$ & $\mu^\mathrm{e}_a$ & $\mu^\mathrm{e}_b$ & $\mu^\mathrm{e}_{c*}$ \\
$\mu_\mathrm{Boltz}$ &              1.198 &              2.735 &                 0.239 &              0.667 &              0.313 &                 0.209 \\
$\mu(\varepsilon)$   &              1.213 &              2.769 &                 0.241 &              0.663 &              0.323 &                 0.209 \\
relative error       &              0.013 &              0.012 &                 0.008 &              0.005 &              0.033 &                 0.003 \\
\end{tabular}
\end{ruledtabular}
\end{table}

\begin{figure}[H]
\centering
  \includegraphics[]
        {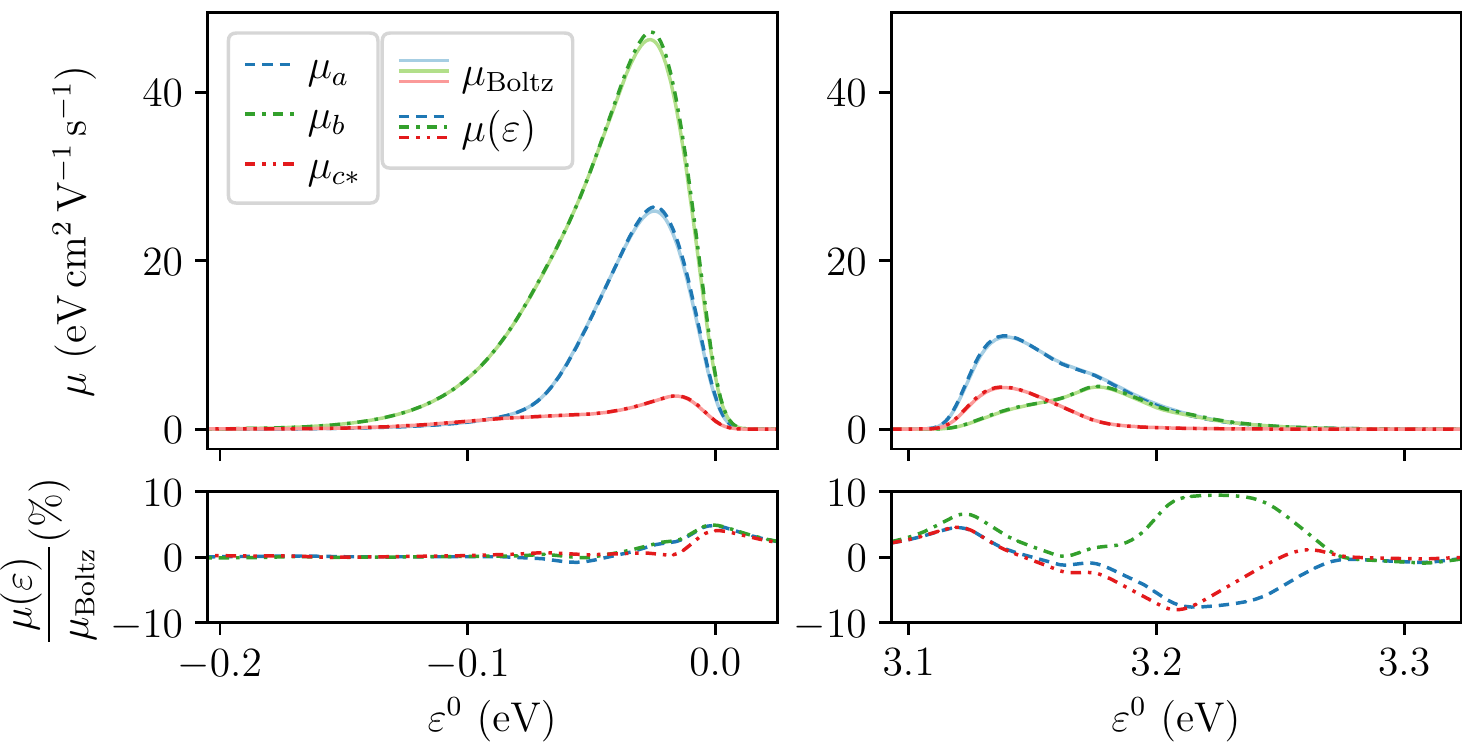}
  \caption{\label{SI:fig:muapprox}%
  Comparing the integrand of the mobility from the explicit summation over $n\vk$ in the self-energy relaxation time approximation in \cref*{eq:muboltz} ($\mu_\mathrm{Boltz}$, solid transparent line) with the product of the four independent functions $D(\varepsilon)$, $f'(\varepsilon)$, $v_{\alpha}^2(\varepsilon)$, and $\tau(\varepsilon)$ from \cref*{eq:muapprox} ($\mu(\varepsilon)$, dashed lines).
  }
\end{figure}

\section{Eigenvalue self-consistent self-energy}

\begin{figure}[H]
\centering
  \includegraphics[]{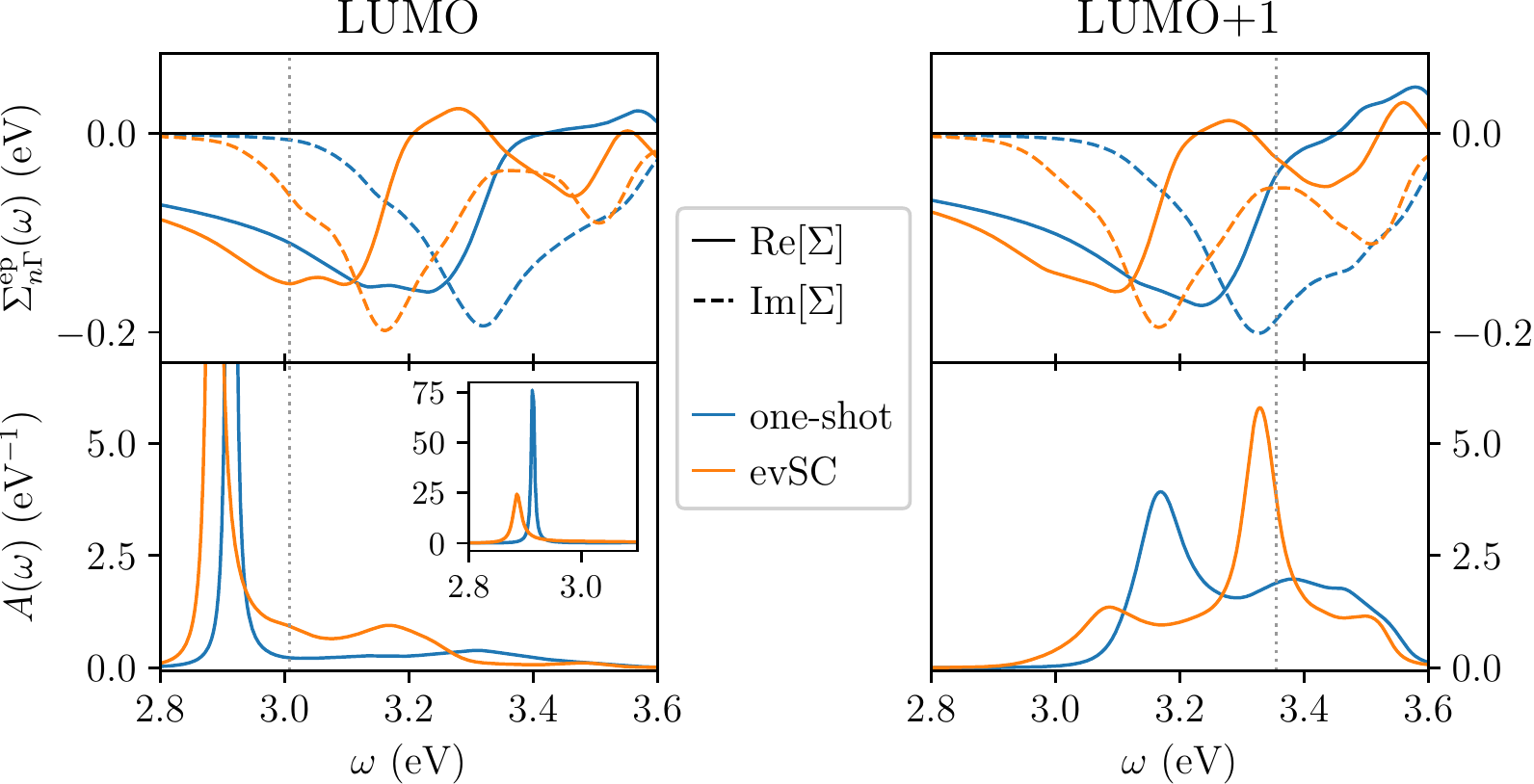}
  \caption{%
    Comparing the self-energyies (top) and spectral functions (bottom) of the one-shot with the eigenvalue self-consistent (evSC) self-energy calculation for the lowest (LUMO) and second lowest (LUMO+1) conduction band at $\Gamma$.
    The DFT eigenvalues are marked with vertical gray dotted lines.
  }\label{SI:fig:se_sf_sc}
\end{figure}

In \cref{SI:fig:se_sf_sc} we compare the eigenvalue self-consistent (evSC) with the one-shot approach by looking at the self-energies and spectral functions of the two lowest conduction bands (LUMO and LUMO+1) at $\Gamma$.
The differences of the frequency-dependent self-energies between the one-shot and self-consistent calculations are quite significant.
The main peaks of the real and imaginary parts are red-shifted by almost 0.2~eV, and a second peak develops at around 3.5~eV, above which the self-energy is mostly unchanged in comparison.
This directly correlates with the renormalized density of states (DOS), where the lower energy states are renormalized by 0.1-0.2~eV, and the higher states remain unshifted.

While the frequency-dependent self-energies of the LUMO and LUMO+1 states look very similar, both for one-shot and evSC, the effect of the self-consistent approach on their spectral functions is quite different.
The quasiparticle peak of the conduction band minimum is slightly red-shifted, and broadened compared to the one-shot.
Additionally, the satellites are more pronounced and also red-shifted.
%
%
Overall, the qualitative shape of the quasiparticle peak remains the same, however.

For the second conduction band LUMO+1, both the position of the main quasiparticle peak as well as the shape of the spectral function are severely altered.
The electronic energy of this state falls into a region, where the evSC electron-phonon self-energy is very non-linear, and thus departs from the textbook quasiparticle picture.
The renormalization of the main quasiparticle peak changes from \SI{-0.17}{eV} for the one-shot calculation to \SI{-0.03}{eV} for evSC.
Furthermore, instead of the one broad phonon satellite in the one-shot calculation, the evSC spectral function shows two satellites, one at higher and one at lower energies.
At other $\vk$-points, the satellite at \SI{3.1}{eV} actually becomes the main quasiparticle peak (not shown here).
A similar transition can be more clearly seen for the valence band in \cref{fig:sf_full_bs}, where about half-way between $\Gamma$ and Z, the quasiparticle peak shifts its weight from \SI{0.3}{eV} to \SI{0}{eV}.
This transition, or seeming quasiparticle ``discontinuity'' in the conduction bands is less visible because the bands are much more broadened out.

\begin{figure}[H]
\centering
    \includegraphics[]{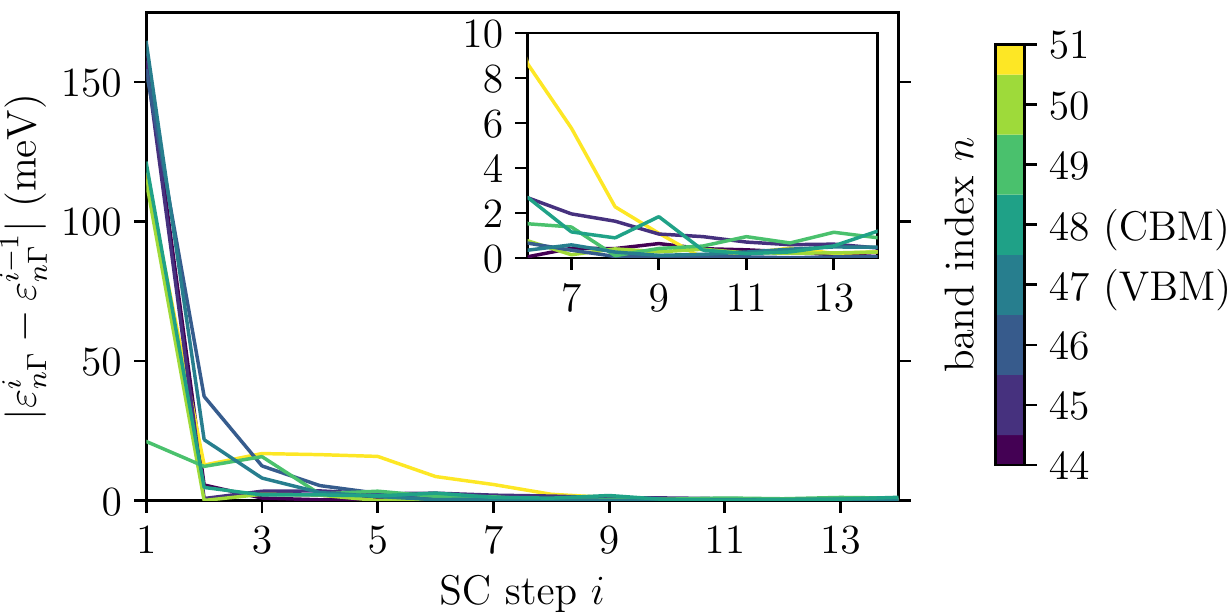}
    \caption{\label{SI:fig:sc_conv}%
    Convergence of the electronic energies using the eigenvalue self-consistent electron-phonon approach for bands around the band gap.
    The energy difference at the first step is equal to the renormalization obtained from the on-the-mass-shell approximation.
    }
\end{figure}


%

\begin{figure}[H]
  \centering
  \includegraphics[]{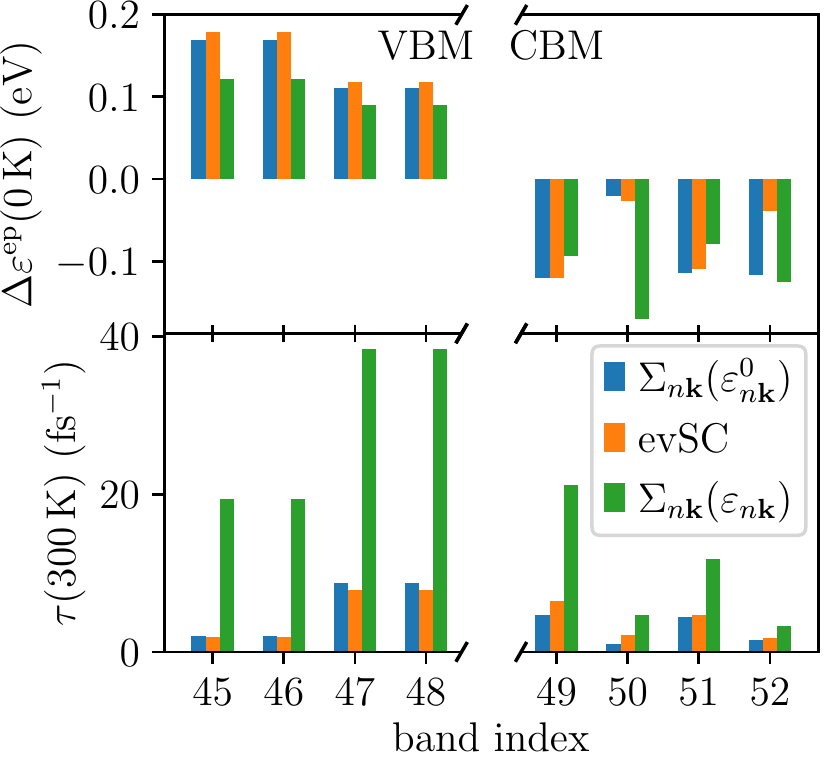}
  \caption{\label{SI:fig:on_off_SC}%
  Comparing the renormalizations and lifetimes of the evSC calculation with the on-the-mass-shell approximation ($\Upsigma(\varepsilon^0)$) and the quasiparticle solution ($\Upsigma(\varepsilon)$). Valence bands are at $\vk=\mathrm{A}$, and conduction bands at $\Gamma$, coinciding with the VBM and CBM, respectively.
  The renormalizations (top) are calculated at \SI{0}{K} using \Vdft{} lattice parameters.
  For the lifetimes (bottom) at \SI{300}{K} we used \Vexp{} to account for lattice expansion.
  Overall, the on-the-mass-shell approximation agrees better with the evSC approach.
}
\end{figure}

\begin{figure}[H]
  \includegraphics[]{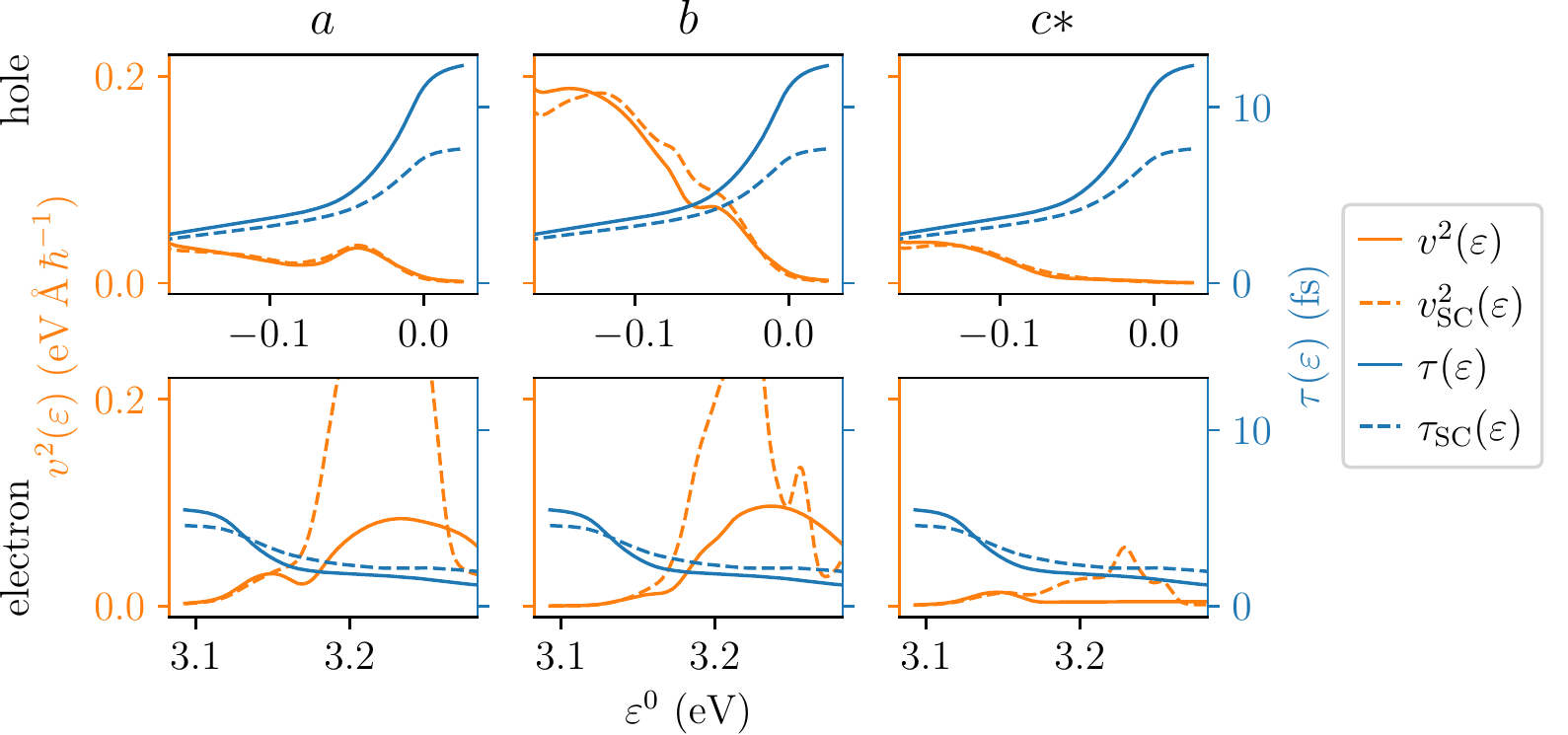}
  \caption{\label{SI:fig:mobilities_decomposition_SC}%
  Comparing the energy dependent velocities (\cref*{eq:v2approx}) and lifetimes (\cref*{eq:tauapprox}) of the hole (top) and electron (bottom) carriers at \SI{300}{K} (\Vdft{}) between the one-shot (solid) and SC (dashed) calculations.
  The high electron velocities around \SI{3.2}{eV} are an artifact caused by the breakdown of the quasiparticle picture near the band splitting.
}
\end{figure}

\bibliography{main}

\makeatletter\@input{ms_aux.tex}\makeatother